\documentclass[sigconf]{acmart}
\usepackage{booktabs} 
\usepackage{graphicx}  
\usepackage{amssymb}
\usepackage{bm}
\usepackage{amsmath}
\usepackage{verbatim}
\usepackage[ruled,vlined]{algorithm2e}
\usepackage{subfigure}
\usepackage{subfigure}
\usepackage{natbib}
\usepackage[Symbol]{upgreek}
\usepackage{tabularx}
\usepackage{multirow}
\usepackage{stfloats}
\usepackage{url}
\usepackage{diagbox}

\setcopyright{rightsretained}

\begin{document}

\copyrightyear{2019}
 \acmYear{2019} 
\setcopyright{iw3c2w3}
\acmConference[WWW '19]{Proceedings of the 2019 World Wide Web Conference}{May 13--17, 2019}{San Francisco, CA, USA}
\acmBooktitle{Proceedings of the 2019 World Wide Web Conference (WWW '19), May 13--17, 2019, San Francisco, CA, USA}
\acmPrice{}
\acmDOI{10.1145/3308558.3313478}
\acmISBN{978-1-4503-6674-8/19/05}
\fancyhead{}

\title{Spectrum-enhanced Pairwise Learning to Rank}

\author{Wenhui Yu}
\affiliation{%
  \institution{Tsinghua University}
  \city{Beijing}
  \state{China}
}
\email{yuwh16@mails.tsinghua.edu.cn}

\author{Zheng Qin}
\authornote{The corresponding author.}
\affiliation{%
  \institution{Tsinghua University}
  \city{Beijing}
  \state{China}
}
\email{qingzh@mail.tsinghua.edu.cn}


\begin{abstract}

To enhance the performance of the recommender system, side information is extensively explored with various features (e.g., visual features and textual features). However, there are some demerits of side information: (1) the extra data is not always available in all recommendation tasks; (2) it is only for items, there is seldom high-level feature describing users. To address these gaps, we introduce the \textit{spectral features} extracted from two hypergraph structures of the purchase records. Spectral features describe the \textit{similarity} of users/items in the graph space, which is critical for recommendation. We leverage spectral features to model the users' preference and items' properties by incorporating them into a Matrix Factorization (MF) model.

In addition to modeling, we also use spectral features to optimize. Bayesian Personalized Ranking (BPR) is extensively leveraged to optimize models in \textit{implicit feedback} data. However, in BPR, all missing values are regarded as negative samples equally while many of them are indeed unseen positive ones. We enrich the positive samples by calculating the similarity among users/items by the spectral features. The key ideas are: (1) similar users shall have similar preference on the same item; (2) a user shall have similar perception on similar items. Extensive experiments on two real-world datasets demonstrate the usefulness of the spectral features and the effectiveness of our spectrum-enhanced pairwise optimization. Our models outperform several state-of-the-art models significantly.

\end{abstract}

%
%

\ccsdesc[500]{Information systems~Collaborative filtering}
\ccsdesc[500]{Information systems~Social recommendation}
\ccsdesc[500]{Information systems~Recommender systems}
\ccsdesc[300]{Human-centered computing~Social recommendation}

\keywords{
Collaborative filtering, spectral feature, pairwise learning to rank, latent community, latent category.}

\maketitle





\section{Introduction}
Recommender systems have been widely used in online services such as E-commerce and social media sites to predict users' preference based on their interaction histories. Modern recommender systems uncover the underlying latent factors that encode the preference of users and properties of items. In recent years, to strengthen the presentation ability of the models, various features are incorporated for additional information, such as visual features from product images \cite{VBPR,matrix,AES}, textual features from review data \cite{Key_Frame,he_context}, and auditory features from music \cite{music}. However, these features are not generally applicable, for example, visual features can only be used in product recommendation, while not in music recommendation. Also, these features are unavailable in some recommendation tasks, such as point-of-interest recommendation \cite{POI1,POI2}. Even they are available, we need extra efforts to collect and process them. Moreover, we can only get features for items while there is seldom high-level feature for users. To address these gaps, we introduce spectral features extracted from the purchase records for both items and users in this paper. We then use the spectral features to (1) model users' preference and items' properties and to (2) optimize the proposed model.

Recommender systems predict the missing value by uncovering the similarity of users/items, thus the information of similarity is vital in recommendation tasks. For example, in user-/item-based collaborative filtering, we recommend by calculating the similarity among users/items \cite{user-based,item-based}; in model-based collaborative filtering, the low-rank form ensures the linear dependence of latent factors, i.e., the similarity among users/items \cite{MF,MF0,FM}. Inspired by this, we propose a new feature that contains the similarity information. We define the front $K$ eigenvectors of a Laplacian matrix as the spectral feature. Devised for spectral clustering \cite{spec_clus}, the spectral feature describes the distance among vertices in a graph space thus contains abundant information of similarity. The spectral feature extracts information from the purchase history rather than additional data, thus it is generally applicable and works in almost all situations. Also, we can extract spectral features for both users and items.

We define the set of users near in the graph space as a \textit{latent community}, which means they have similar purchase behaviors. For an item preferred by a specific user, it may get a high score for her latent community members. Also, the set of items near in the graph space is defined as a \textit{latent category}, items in the same latent category have similar properties. Users who like certain item may have interests in the latent category the item belongs to. Of special notice is that a latent category is not like a real category, items in it may be relevant items, similar items, etc., in a word, items strongly connected in the graph. In this paper, we incorporate the spectral features into a Matrix Factorization (MF) model \cite{BPR,MF} to propose our \textbf{S}pectrum-enhanced \textbf{C}ollaborative \textbf{F}iltering (\textbf{SCF}) model. By learning spectral dimensions, SCF uncovers users' preference for latent categories and items' fitness for latent communities.

Compared with other side information features, our spectral feature is more suitable for the \textit{implicit feedback} data. Implicit feedback data is like ``purchase'' or ``browse'' in E-commerce sites, ``like'' in social media sites, ``click'' in advertisements, etc. In real-world applications, the data of user behaviours in ``one-class'' (implicit feedback) is easier to collect and more generally applicable than ``multi-class'' scores (\textit{explicit feedback}). The spectral features, which are independent of extra data, are more suitable for the implicit feedback data, since they maintain the advantages of easy to collect and generality.

Besides providing information of similarity, we also use the spectral features to enhance the pairwise learning. When optimizing model on implicit feedback data, Bayesian Personalized Ranking (BPR) is widely used due to the outstanding performance \cite{BPR,VBPR,AES}. It aims to maximize the likelihood of pairwise preference over positive samples and negative samples. However, there is a critical issue: All missing values are simply treated as negative samples in BPR. In fact, some of the missing values are indeed unlabelled positive samples, users may like them but just have not seen them yet. To deal with this issue, we cluster all items and users by spectral features to construct the latent categories and latent communities, respectively. We assume that a user shows stronger preference for items that are in the same latent category with what she purchased, or items that purchased by her neighbors in the same latent community, than other missing entries. Considering the enriched preference relationship, we propose an optimization method called \textbf{S}pectrum-enhanced \textbf{P}airwise \textbf{L}earning to \textbf{R}ank (\textbf{SPLR}), and optimize SCF with it. Finally, we validate the effectiveness of our proposed model by comparing it with several baselines on the \emph{Amazon.Clothes} and \emph{Amazon.Jewelry} datasets. Extensive experiments show that we improve the performance significantly by exploring spectral features.

Specifically, our main contributions are listed as follows:

\begin{itemize}
\item{We leverage novel spectral features in recommendation tasks to capture the similarity information of users/items, and propose an SCF model by injecting these features to the MF model.}

\item{We propose a spectral clustering-enhanced pairwise ranking method, SPLR, to optimize our model. We construct latent categories and latent communities to enrich the positive samples.}

\item{We extend our model and propose a framework to use side information features for modeling and for optimization. We can explore any kinds of features in our framework, including spectral features and other conventional features. Multiple features can be utilized at a time.}

\item{We devise comprehensive experiments on two real-world datasets to demonstrate the effectiveness of our proposed methods.}
\end{itemize}

\section{Related Work}
After Matrix Factorization (MF) is utilized to deal with recommendation tasks \cite{MF0,MF,BPR}, modern recommender systems develop rapidly. MF models learn the latent factors of users and items by reconstructing the purchase records in a low-rank form. Latent factors represent the preference of users and properties of items. Many variants are proposed to promote the ability to model the complex preference that users exhibit toward items based on their past interactions. \cite{NCF,H1} used deep structure to learn embedding. \cite{he_Attentive,cc} took time into account when making predictions. \cite{O2,DCF} focused on fast algorithms for online recommendation.

\subsection{Side Information Features}
One important way to enhance the presentation ability is to leverage the side information. Theoretically, latent dimensions can capture all relevant factors, but they usually cannot in applications due to the sparsity of the datasets, thus extra information is desired. The visual features are widely used since users' decisions depend largely on products' appearance \cite{VBPR,Image_based,AES,matrix}. \cite{VBPR,Image_based} predicted consumers' behavior with the CNN feature. \citet{AES} utilized the aesthetic feature to model users' aesthetic preference on clothes. \citet{matrix} leveraged several visual features to recommend movies. There are also many efforts exploring the textural feature to recommend \cite{text1,Key_Frame}, \citet{text1} proposed models with the textural feature from the review data and \citet{Key_Frame} extracted the textural feature from time-synchronized comments for key-frame recommendation. In music recommendation, the auditory feature is generally used, \citet{music} extracted the auditory feature by a pre-trained deep structure to recommend music. 

Various features are used for different kinds of information. However, one feature, the spectral feature, has never been explored. In this paper, we extract the spectral features from the hypergraphs of users and items, which represent the similarity of vertices in the hypergraphs. Compared with features mentioned above, the spectral features are independent of additional information so they are suitable for more situations. Moreover, conventional features are all for items, while we extract spectral features for both users and items.

\subsection{Graph-enhanced Recommendation}
In recent years, hypergraph gains increasing attention in the recommendation domain \cite{hyper,hyper_nips,SR1}. \citet{hyper} proposed a user-based collaborative filtering enhanced with the hypergraph: The similarity of the users is calculated with the hypergraph embedding. \cite{hyper_nips,SR1} filtered the latent factors with hypergraph regularized term to smooth them. There are also some efforts promoting the recommendation performance with social networks \cite{social,trust_pro}. \citet{social} calculated similarities with the social networks embedding for memory-based collaborative filtering. \citet{trust_pro} calculated the latent factors of a user with its neighbors to give the prediction. Graph convolution networks are also widely used in recommendation tasks \cite{graph_convolution,Spectral_CF}. \citet{graph_convolution} proposed a graph convolutional auto-encoder that learns the structural information of a graph for latent factors of users and items. \citet{Spectral_CF} constructed random walk laplacian matrix of the user-item bipartite graph and then introduced a deep spectral convolutional network to capture the user-item connectivity information.

In this paper, we introduce a new way to explore graph structures in recommendation tasks. We extract features from the Laplacian matrix of the hypergraphs of users and items, and make prediction with the MF term jointly. We also utilize these features to optimize our proposed method.

\subsection{Learning to Rank}
As implicit feedback data is easier to collect, it is extensively used in real-world application. However, prediction on implicit feedback dataset is a challenging task because there are only positive samples and unobserved samples. We cannot discriminate negative samples and unlabeled positive samples from the unobserved ones. \citet{BPR} treated all unobserved samples as negative ones when sampling while some of them are indeed unlabelled positive samples. Users may like them but just have not seen them yet. To address this gap, many works improved the pairwise learning to rank method. It is assumed that all users are independent in BPR, \citet{groupBPR} tried to relax this constraint and proposed a method called group preference-based Bayesian personalized ranking (GBPR), which modeled the preference of user groups. \citet{itemgroupBPR} constructed the preference chain of item groups for each user. \citet{CPLR} utilized collaborative information mined from the interactions between users and items. \cite{dynamicBPR,improve_pairwise} proposed dynamic negative sampling strategies to maximize the utility of a gradient step by choosing ``difficult'' negative samples. \cite{viewBPR,adaptiveBPR} used view information to enrich positive samples. \cite{listrank,Listwise} proposed listwise ranking methods instead of pairwise ones. \citet{uninteresting} utilized both implicit and explicit feedback data to improve the quality of negative sampling. 

In existing efforts, only low-order connections are considered when measuring the similarity among vertices \cite{groupBPR,itemgroupBPR,CPLR}. In this paper, we use the spectral features, which contains the information of high-order connections, to enhance the pairwise learning. We cluster all items/users by spectral features to construct latent categories/communities. For vertices (users/items) in the same cluster, they are strongly connected\footnote{We use ``strongly connected'' to indicate that vertices are connected by many paths in the graph (including direct connections or high-order connections), which is different from the conception in directed graphs \cite{strong-connected}.} in the hypergraph thus are very similar to each other. For each user, we regard items in the same latent category with her positive samples and items purchased by her latent community members as the potential samples, and assume that the user prefers them than other negative samples.

\section{Spectrum-enhanced Collaborative Filtering}
In this section, we propose a novel recommendation model called Spectrum-enhanced Collaborative Filtering (SCF). We first introduce the spectral features and then inject them into an MF model. In this paper, bold uppercase letters refer to matrices. For example, $\bm{{{\rm A}}}$ is a matrix, $\bm{{{\rm A}}}_i$ is the $i$-th row of $\bm{{{\rm A}}}$, $\bm{{{\rm A}}}_{*,j}$ is the $j$-th column of $\bm{{{\rm A}}}$, and $\bm{{{\rm A}}}_{ij}$ is the value at the $i$-th row and the $j$-th column of $\bm{{{\rm A}}}$. $\bm{{{\rm A}}}^U$ is the matrix for user and $\bm{{{\rm A}}}^I$ is the matrix for item, $\bm{{{\rm A}}}^{(k)}$ is the $k$-th matrix.

\subsection{Spectral Feature}
In a recommendation task, we use matrix $\bm{{{\rm R}}} \in \mathbb{R}^{N \times M}$ to denote the interactions between users and items (there are $N$ users and $M$ items in total). $\bm{{{\rm R}}}_{ui} = 1$ if user $u$ purchased items $i$ and $\bm{{{\rm R}}}_{ui} = 0$ otherwise. Our task is to predict the missing values (0 in $\bm{{{\rm R}}}$) to recommend top-$n$ items for each user.

The hypergraph is a generalization of the simple graph, where an hyperedge (edge in hypergraph) connects any number of vertices rather than just two. A hypergraph is usually represented with the incidence matrix $\bm{{{\rm H}}}\in \mathbb{R}^{N \times M}$ (there are $N$ vertices and $M$ hyperedges). Each row in $\bm{{{\rm H}}}$ is for a vertex and each column is for a hyperedge. $\bm{{{\rm H}}}_{ij} = 1$ if vertex $i$ is connected by hyperedge $j$ and $\bm{{{\rm H}}}_{ij} = 0$ otherwise. We can see that a hypergraph can represent the purchase records by nature. We define two hypergraph structures, a user hypergraph and an item hypergraph. In the user hypergraph, users are vertices and items are hyperedges while in the item hypergraph, items are vertices and users are hyperedges.

\begin{figure}[ht!]
    \setlength{\abovecaptionskip}{2mm}
    \centering
    \subfigure[Interaction Matrix]{
        \includegraphics[scale = 0.35]{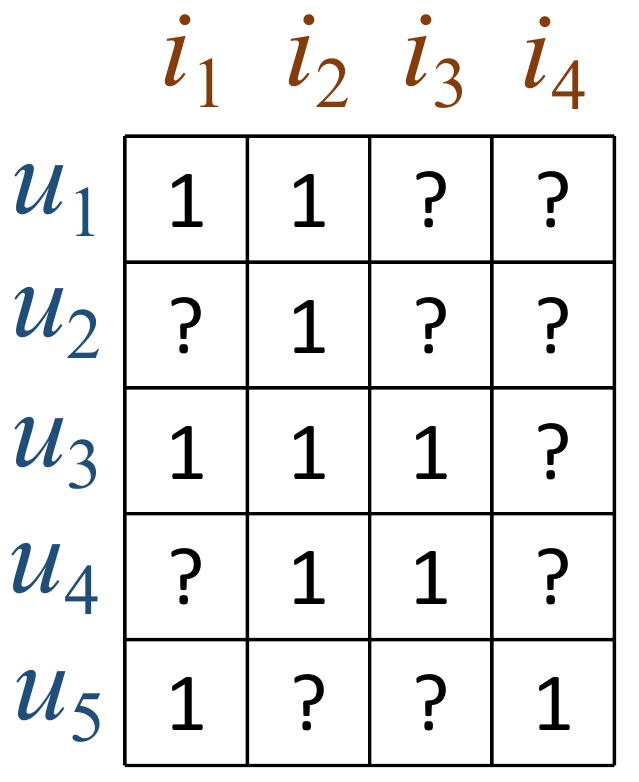}
        \label{subfig:inter}
    }
    \hspace{-1mm}
    \subfigure[User Hypergraph]{
        \includegraphics[scale = 0.3]{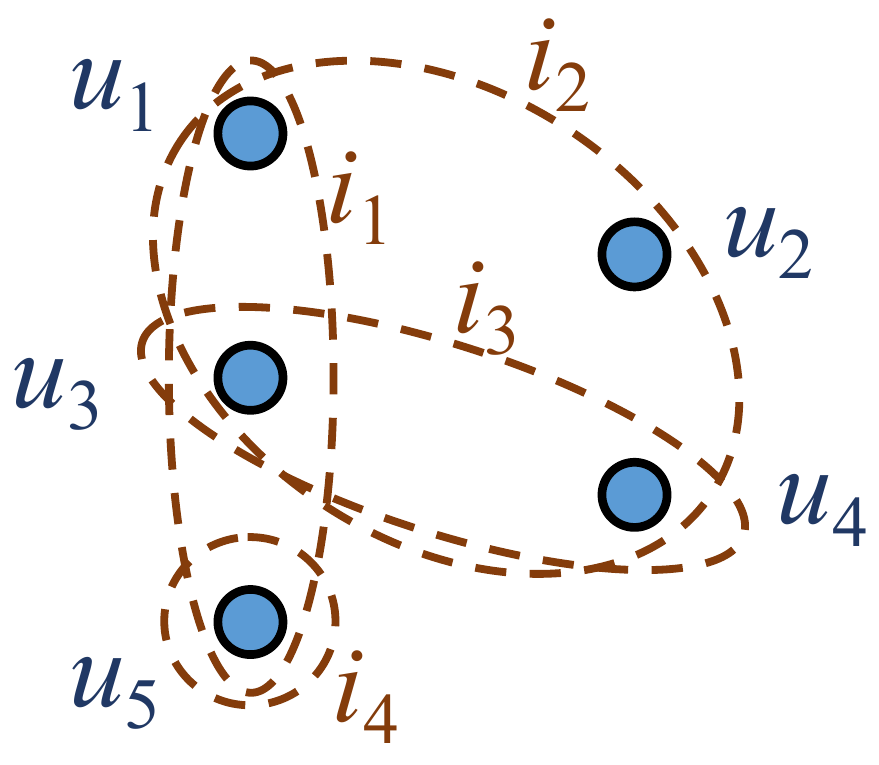}
        \label{subfig:hyper_u}
    }
    \hspace{-1mm}
    \subfigure[Item Hypergraph]{
        \includegraphics[scale = 0.3]{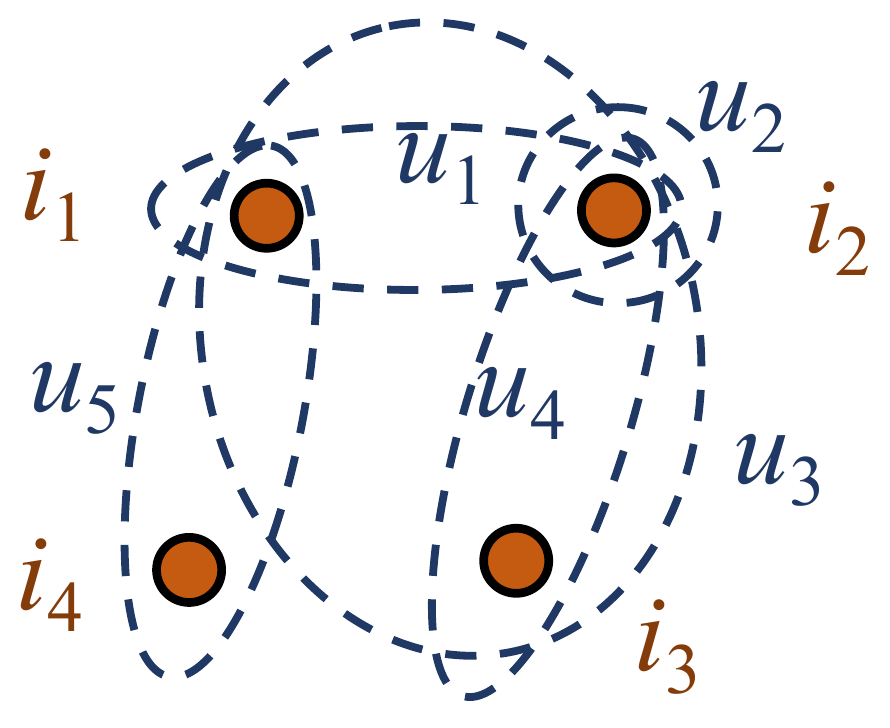}
        \label{subfig:hyper_i}
    }
    \caption{An example for user hypergraph and item hypergraph. For the user hypergraph, the incidence matrix $\bm{{{\rm H}}}^U={\bm{{\rm R}}}$ and for the item hypergraph, $\bm{{{\rm H}}}^I={\bm{{\rm R}}}^\mathsf{T}$.}
    \label{fig:hypers}
\end{figure}

For a hypergraph represented with the incidence matrix $\bm{{{\rm H}}}$, the Laplacian matrix $\bm{{{\rm L}}}$ is defined as \cite{hyper_laplacian}:
\begin{flalign}
\label{equ:laplacian}
\bm{{{\rm L}}} = {\bm{{\rm D}}}^{-\frac{1}{2}} ({\bm{{\rm D}}} - {\bm{{\rm H}}} {\bm{{\rm W}}} {\bm{{\rm \Delta}}}^{-1} {\bm{{\rm H}}}^\mathsf{T})  {\bm{{\rm D}}}^{-\frac{1}{2}},
\end{flalign}
where the diagonal matrix $\bm{{{\rm D}}}\in \mathbb{R}^{N \times N}$ denotes the degrees of vertices, the diagonal matrix $\bm{{{\rm W}}} \in \mathbb{R}^{M \times M}$ denotes the weights of hyperedges, and the diagonal matrix $\bm{{{\rm \Delta}}}\in \mathbb{R}^{M \times M}$ denotes the degrees of hyperedges. Laplacian matrix is a difference operator of a hypergraph \cite{gra_conv}, for any signal $\bm{{{\rm S}}}\in \mathbb{R}^{N \times L}$, it satisfies: $$\left({\bm{{\rm L}}} {\bm{{\rm S}}}\right)_i = \sum_{j\in \mathcal{N}_i} \sum_{k=1}^M {\bm{{\rm H}}}_{ik} \frac{{\bm{{\rm W}}}_{kk}}{{\bm{{\rm \Delta}}}_{kk}} {\bm{{\rm H}}}_{jk}\left(\frac{{\bm{{\rm S}}}_i}{\sqrt{{\bm{{\rm D}}}_{ii}}} - \frac{{\bm{{\rm S}}}_j}{\sqrt{{\bm{{\rm D}}}_{jj}}} \right),$$ where $\mathcal{N}_i$ is the set of vertices connected to vertex i. We can see the effect of the Laplacian matrix is to take the first-order difference in the neighborhood of vertex $i$. In many machine learning tasks, the parameter $\bm{{{\rm S}}}$ is smoothed by minimizing the Laplacian regularization term $trace({\bm{{\rm S}}}^\mathsf{T}{\bm{{\rm L}}}{\bm{{\rm S}}})$ \cite{hyper_nips,He_hyper}.

To extract the spectral feature, we factorize the Laplacian matrix with eigen-decomposition: $\bm{{{\rm L}}} = {\bm{{\rm \Phi}}} {\bm{{\rm \Lambda}}} {\bm{{\rm \Phi}}}^\mathsf{T}$, where $\bm{{{\rm \Lambda}}} = diag(\lambda_1,$ $\lambda_2,$ $\cdots,$ $\lambda_N)$ is the eigenvalue matrix, all eigenvalues are in ascending order, i.e., $\lambda_1 \leq \lambda_2\leq\cdots\leq\lambda_N$. We define the matrix formed by the first $K$ eigenvectors as the spectral feature matrix: $\bm{{{\rm F}}} = {\bm{{\rm \Phi}}}_{*,1:K}$,
and $\bm{{{\rm F}}}_i$ is the spectral feature of vertex $i$. The spectral feature can be used for spectral clustering \cite{spec_clus,hyper} and graph Fourier transform \cite{gra_conv}. It contains the information of the similarity among vertices thus we can measure the similarity of two vertices $i$ and $j$ with ${\bm{{\rm F}}}_i{\bm{{\rm F}}}_j^\mathsf{T}$. Though eigen-decomposition is computationally expensive (taking the user hypergraph as an example, with $O(N^3)$ time complexity, where $N$ is the user number), however, $\bm{{{\rm L}}}$ is highly sparse in recommendation tasks. Considering there are $O(N)$ non-zero elements in $\bm{{{\rm L}}}$, the time complexity could be $O(K^2N)$ with Lanczos method \cite{lanczos1,lanczos}, which is much smaller than $O(N^3)$.

\subsection{Hybrid Model}
\label{sec:hybrid_model}
For the hybrid model (SCF), we incorporate the spectral features into a basic MF model, which is the state-of-the-art for rating prediction as well as modeling implicit feedback, to reconstruct the interaction matrix with the low-rank form:
\begin{flalign}
\label{equ:hybrid_model}
\hat{ {\bm{{\rm R}}} } = {\bm{{\rm U}}} {\bm{{\rm V}}}^\mathsf{T} + {\bm{{\rm P}}} {\bm{{\rm F}}}^\mathsf{T} + {\bm{{\rm E}}} {\bm{{\rm Q}}}^\mathsf{T},
\end{flalign}
where $\hat{ {\bm{{\rm R}}} } \in \mathbb{R}^{N \times M}$ is the reconstruction, $\bm{{{\rm U}}} \in \mathbb{R}^{N \times K_0}$ and $\bm{{{\rm V}}} \in \mathbb{R}^{M \times K_0}$ are the latent factors of users and items respectively. $\bm{{{\rm U}}}_i$ is the preference of user $i$ and $\bm{{{\rm V}}}_j$ is the properties of item $j$. $\bm{{{\rm E}}} \in \mathbb{R}^{N \times K_1}$ and $\bm{{{\rm F}}} \in \mathbb{R}^{M \times K_2}$ are the spectral features of users and items respectively. $\bm{{{\rm P}}} \in \mathbb{R}^{N \times K_2}$ is the preference matrix of users, $\bm{{{\rm Q}}} \in \mathbb{R}^{M \times K_1}$ is the fitness matrix of items.

Considering that strongly connected users/items shall have similar preference/properties, $\bm{{{\rm E}}}$/$\bm{{{\rm F}}}$ contains the information of similarity among users/items. We use the similarity information to enhance the MF predictor. We call the three terms in Equation (\ref{equ:hybrid_model}) as model-based, item-based, and user-based collaborative terms respectively. In a conventional item-based collaborative filtering model, we calculate the similarity between each pair of items. For a positive sample $i$ and a missing sample $j$, if $j$ is similar with $i$, we recommend $j$ to current user. Similarly, in the second term of our predictor, for current user $u$ and her purchased item $i$, $\bm{{{\rm P}}}_u {\bm{{\rm F}}}^\mathsf{T}_i$ is high. For a similar item $j$, ${\bm{{\rm F}}}_i{\bm{{\rm F}}}_j^\mathsf{T}$ is high, thus $\bm{{{\rm P}}}_u {\bm{{\rm F}}}^\mathsf{T}_j$ is high as well, and $j$ can be recommended to $u$. We can see that this procedure is just the same as item-based collaborative filtering, so we call the second term of Equation (\ref{equ:hybrid_model}) item-based collaborative term. And the third term, for the same reason, is called user-based collaborative term.

Now we briefly discuss the advantage of our item-based collaborative term over conventional item-based collaborative filtering. In item-based collaborative filtering, when calculating the similarity between a pair of items, only first-order connections are taken into account, while in our item-based collaborative term, high-order connections are also considered. Take Figure \ref{subfig:hyper_i} as an example, $i_3$ and $i_4$ are not connected by certain hyperedge directly, hence the similarity is 0 in conventional item-based collaborative filtering. While in our model, connections $i_4$-$i_1$-$i_3$ and $i_4$-$i_1$-$i_2$-$i_3$ are also considered, therefore $sim(i_3,i_4)>0$ (we will discuss the reason detailedly in the next section). In fact, in Equation (\ref{equ:hybrid_model}), three models give the final prediction jointly: An MF model, and two hypergraph spectrum-enhanced memory-based collaborative filtering models.
 
\section{Spectrum-enhanced Pairwise Learning}
Besides providing information to model users' preference and items' property, spectral features are also used to optimize the model. Optimization on implicit feedback data usually takes the form of pairwise learning, which maximizes the likelihood of relative preference over a pair of positive and negative feedbacks:
\begin{flalign}
{\rm BPR\_O{\scriptstyle PT}} = \sum_{u\in\mathcal{U}} \sum_{i\in \mathcal{I}^+_u} \sum_{j\in \mathcal{I}\setminus\mathcal{I}^+_u} {\rm ln} \,\sigma\! \left(\hat{{\bm{{\rm R}}}}_{uij}\right) - \frac{\lambda_r}{2} {\left\Arrowvert {\bm{{\rm \Theta}}} \right\Arrowvert}_{\rm F}^2\, . \nonumber
\end{flalign}
where $\mathcal{U}$ and $\mathcal{I}$ are the sets of users and items. $\mathcal{I}_u^+$ is the set of positive items of $u$. $\sigma(\;)$ is the sigmoid function. $\hat{\bm{{\rm R}}}$ is defined in the Equation (\ref{equ:hybrid_model}) and $\Hat{{\bm{{\rm R}}}}_{uij} = \Hat{{\bm{{\rm R}}}}_{ui} - \Hat{{\bm{{\rm R}}}}_{uj}$. The last term is the regularization term to prevent overfitting, where $\lambda_r$ is the regularization coefficient, ${\left\Arrowvert \;\; \right\Arrowvert}_{\rm F}$ is the Frobenius norm of the matrix, and $\bm{{{\rm \Theta}}}$ represents the parameters of the model.

There is a critical issue: A user did not purchase an item may not because she has no interest in it, but just because she has never seen it yet. Our task is to uncover users' preference and recommend them unseen items they are interested in. However, in BPR, all missing entries are treated as negative samples nevertheless some of them are indeed unlabeled positive samples. To address this gap, we enrich the positive samples by the spectral clustering.

\subsection{Objective Function}
We cluster all vertices (items/users) by the normalized spectral features. Vertices in the same cluster are strongly connected in the hypergraph, though may not be connected directly. 

\noindent\textbf{Latent category:} We define a cluster of items as a latent category, items in it are of the similar kind, or highly relevant, since they are purchased by the same user or users with similar preference. For certain item $i$, the latent category it belongs to is denoted as $\mathcal{C}_i$. We argue that if a user likes $i$, she may like the items in $\mathcal{C}_i$ with a high probability. 

\noindent\textbf{Latent community:} Similarly, we define a cluster of users as a latent community, users in it purchase the same items, similar items, or relevant items, thus they have similar preference. For certain user $u$, the latent community she belongs to is denoted as $\mathcal{C}_u$. If a user likes $i$, her latent community members may like $i$ as well.

To give the relative preference, we construct three sets for each user $u$:
\begin{flalign}
\label{equ:three_sets}
\left\{
\begin{array}{ll}
\mathcal{I}^+_u = \{i | {\bm{{\rm R}}}_{ui} = 1\}, &\text{Positive set}\\
\mathcal{P}_u = \mathcal{P}_u^U \bigcup \mathcal{P}_u^I, &\text{Potential set}\\
\mathcal{I}^-_u = \mathcal{I} - \mathcal{I}^+_u - \mathcal{P}_u, &\text{Negative set}\\
\end{array},
\right.
\end{flalign}
where $\mathcal{P}_u^U = \{i | i\in \mathcal{I}^+_v, v \in \mathcal{C}_u \& i\notin \mathcal{I}^+_u\}$ is the user-based collaborative potential set,  $\mathcal{P}_u^I = \{j | j \in \mathcal{C}_i, i\in \mathcal{I}^+_u \& j\notin \mathcal{I}^+_u\}$ is the item-based collaborative potential set.
We have the preference relationship:
$$(u,\mathcal{I}^+_u)\succ(u,\mathcal{I}^-_u),\;(u,\mathcal{I}^+_u)\succ(u,\mathcal{P}_u),\;(u,\mathcal{P}_u)\succ(u,\mathcal{I}^-_u).$$
We can see that, $\mathcal{P}_u$ is the set of items that $u$ may have interests in returned by the memory-based collaborative filtering.

BPR tries to optimize the standard AUC that is designed for binary classification \cite{AUC}, while we try to optimize a generalized AUC (GAUC) \cite{GAUC,CPLR},
\begin{flalign}
{\rm GAUC} = \frac{1}{|\mathcal{U}|} \sum_{u\in \mathcal{U}} \big[&{\rm AUC}(u, \mathcal{I}^+_u, \mathcal{I}^-_u) + \eta_1 {\rm AUC}(u, \mathcal{I}^+_u, \mathcal{P}_u) \nonumber\\
+&\eta_2 {\rm AUC}(u, \mathcal{P}_u, \mathcal{I}^-_u)\big], \nonumber
\end{flalign}where ${\rm AUC}(u, \mathcal{A}, \mathcal{B})=\frac{1}{|\mathcal{A}||\mathcal{B}|}\sum\limits_{i\in \mathcal{A}}\sum\limits_{j\in \mathcal{B}}\delta\left(\hat{{\bm{{\rm R}}}}_{uij}>0\right)$ is the standard AUC, $\eta_1$ and $\eta_2$ are confidence coefficients. We use differentiable loss $\delta(x>0)$ which is identical to $\ln\sigma(x)$. We can see that GAUC is the weighted combination of three standard AUC terms. To maximize it, we propose our \textbf{S}pectrum-enhanced \textbf{P}airwise \textbf{L}earning to \textbf{R}ank (\textbf{SPLR}) optimization:
\begin{flalign}
\label{equ:objective_function}
{\rm SPLR\_O{\scriptstyle PT}}\ = \sum_{u\in \mathcal{U}} \Bigg[& \sum_{i\in \mathcal{I}^+_u} \sum_{j\in \mathcal{I}^-_u} {\rm ln} \,\sigma\! \left(\hat{{\bm{{\rm R}}}}_{uij}\right) \nonumber\\
+\eta_1 &\sum_{i\in \mathcal{I}^+_u} \sum_{j\in \mathcal{P}_u} {\rm ln} \,\sigma\! \left(\hat{{\bm{{\rm R}}}}_{uij}\right) \\
+\eta_2 &\sum_{i\in \mathcal{P}_u} \sum_{j\in \mathcal{I}^-_u} {\rm ln} \,\sigma\! \left(\hat{{\bm{{\rm R}}}}_{uij}\right)\Bigg]
- \frac{\lambda_r}{2} {\left\Arrowvert {\bm{{\rm \Theta}}} \right\Arrowvert}_{\rm F}^2 .\nonumber
\end{flalign}

\subsection{Model Learning}
To maximize the SPLR objective function, we take the first-order derivatives of Equation (\ref{equ:objective_function}) with respect to each model parameter:

\begin{flalign}
\label{equ:derivative1}
{\nabla_{\bm{{\rm \Theta}}}{\rm SPLR\_O\scriptstyle PT}}\ = \sum_{u\in \mathcal{U}} \Bigg[& \sum_{i\in \mathcal{I}^+_u} \sum_{j\in \mathcal{I}^-_u} \sigma \Big(\! -\hat{\bm{{\rm R}}}_{uij} \Big)  \frac{\partial \hat{\bm{{\rm R}}}_{uij}}{\partial {\bm{{\rm \Theta}}}} \nonumber\\
+\eta_1 &\sum_{i\in \mathcal{I}^+_u} \sum_{j\in \mathcal{P}_u} \sigma \Big(\! -\hat{\bm{{\rm R}}}_{uij} \Big)  \frac{\partial \hat{\bm{{\rm R}}}_{uij}}{\partial {\bm{{\rm \Theta}}}}\\
+\eta_2 &\sum_{i\in \mathcal{P}_u} \sum_{j\in \mathcal{I}^-_u} \sigma \Big(\! -\hat{\bm{{\rm R}}}_{uij} \Big)  \frac{\partial \hat{\bm{{\rm R}}}_{uij}}{\partial {\bm{{\rm \Theta}}}} \Bigg]
- \! \lambda_r\, {\bm{{\rm \Theta}}}.\nonumber
\end{flalign}We use $\bm{ \uptheta}$ to denote certain row of $\bm{{{\rm \Theta}}}$, the derivatives in Equation (\ref{equ:derivative1}) are:
\begin{equation}
\label{equ:derivative2}
\frac{\partial \hat{\bm{{\rm R}}}_{uij}}{\partial {\rm \bm \uptheta}} = \left\{
\begin{array}{lcl}
{{\bm{{\rm V}}}_i - {\bm{{\rm V}}}_j} &\text{if} &{\rm \bm \uptheta} = {\bm{{\rm U}}}_u \\
{{\bm{{\rm U}}}_u / -{\bm{{\rm U}}}_u} &\text{if} &{\rm \bm \uptheta} = {\bm{{\rm V}}}_i / {\bm{{\rm V}}}_j \\
{{\bm{{\rm F}}}_i - {\bm{{\rm F}}}_j} &\text{if} &{\rm \bm \uptheta} = {\bm{{\rm P}}}_u\\
{{\bm{{\rm E}}}_u / -{\bm{{\rm E}}}_u} &\text{if} &{\rm \bm \uptheta} = {\bm{{\rm Q}}}_i / {\bm{{\rm Q}}}_j 
\end{array}.
\right.
\end{equation}$\frac{\partial \hat{\bm{{\rm R}}}_{uij}}{\partial {\bm \uptheta}}$ in Equation (\ref{equ:derivative2}) is certain row of $\frac{\partial \hat{\bm{{\rm R}}}_{uij}}{\partial {\bm{{\rm \Theta}}}}$ in Equation (\ref{equ:derivative1}), for example, the $u$-th row when $\bm{ \uptheta} = {\bm{{\rm U}}}_u$.

\begin{algorithm}[ht!]
\caption{Mini-batch gradient descent algorithm}
\label{alg:mSGD}
\LinesNumbered 
\small
\KwIn{Implicit feedback matrix $\bm{{\rm R}}$, regularization coefficient $\lambda_r$, batch size $b$, learning rate $\eta$, weighting parameters $\eta_1$ and $\eta_2$, sampling rate $m$, maximum number of iterations $iter\_max$.}
\KwOut{top-$n$ predictions given by the completed matrix $\hat{\bm{{\rm R}}}$}
calculate the Laplacian matrices of user hypergraph $\bm{{{\rm L}}}^U$ and item hypergraph $\bm{{{\rm L}}}^I$ from $\bm{{\rm R}}$\;
decompose the Laplacian matrices to get the spectral features $\bm{{\rm E}}$ and $\bm{{\rm F}}$\;
cluster vertices and calculate the potential set for each user\;
initialize $\bm{{\rm \Theta}}$ randomly\;
\For{$iter = 1$ to $iter\_max$}{
    split all purchase records into $b$-size batches\;
    \For{each batch}{
    	\For{each record in current batch}{
    	    select $m$ potential items randomly from $\mathcal{P}_u$\;
        	select $m$ negative items randomly from $\mathcal{I}_u^-$\;
            add these items to the current batch\;
        }
        calculate $\nabla_{\bm{{\rm \Theta}}} \rm SPLR\_O\scriptstyle PT$ with current batch\;
        $\bm{{\rm \Theta}} = \bm{{\rm \Theta}}+\eta \nabla_{\bm{{\rm \Theta}}} \rm SPLR\_O\scriptstyle PT$\;   
    }
    calculate $\hat{{\bm{{\rm R}}}}$ to predict the top-$n$ items\;
}
\Return{the top-$n$ items}\;
\end{algorithm}

The detailed procedure is shown in Algorithm \ref{alg:mSGD}. We calculate the Laplacian matrices with the Equation (\ref{equ:hybrid_model}) (line 1) and decompose them, the first $K_1$/$K_2$ eigenvectors of $\bm{{\rm L}}^U$/$\bm{{\rm L}}^I$ form $\bm{{\rm E}}$/$\bm{{\rm F}}$ (line 2). We then construct the three sets in Equation (\ref{equ:three_sets}) for each user (line 3). Lines 5-14 show the process of model learning. For each record $\{u,i\}$, we choose $m$ potential samples $\{j_1,\cdots,j_m\}$ from $\mathcal{P}_u$ (line 9) and $m$ negative samples $\{k_1,\cdots,k_m\}$ from $\mathcal{I}_u^-$ (line 10), $m$ is the sampling rate. The model is optimized with the pairs $\{u,i,j_1\}$, $\cdots$, $\{u,i,j_m\}$, $\{u,i,k_1\}$, $\cdots$, $\{u,i,k_m\}$, $\{u,j_1,k_1\}$, $\cdots$, $\{u,j_m,k_m\}$. We finally calculate the derivatives given by Equation (\ref{equ:derivative1}) with a batch of pairs (line 12) and update the parameters (line 13). 

Like many existing works \cite{socialBPR,groupBPR,itemgroupBPR}, we also use the connections of users and items to enhance the pairwise learning. However, in previous works, only one-order connections are utilized while in our SPLR model, we leverage connections with all orders. In a hypergraph, we have $\bm{{{\rm L}}}^n \!=\! ({\bm{{\rm I}}} \!-\! {\bm{{\rm A}}})^n \!=\! \sum\limits_{r=1}^nC^r_n(-{\bm{{\rm A}}})^r$, where $\bm{{{\rm A}}} \!=\! {\bm{{\rm D}}}^{-\frac{1}{2}} {\bm{{\rm H}}} {\bm{{\rm W}}} {\bm{{\rm \Delta}}}^{-1} {\bm{{\rm H}}}^\mathsf{T} {\bm{{\rm D}}}^{-\frac{1}{2}}$  is the normalized adjacency matrix. $\bm{{{\rm A}}}^r$ indicates the $r$-order connections, therefore $\bm{{{\rm L}}}^n$ contains the information of all connections less than $n$-order in the hypergraph. For the $i$-th eigenvector $\bm{{{\rm u}}}_i$ of $\bm{{{\rm L}}}$, $\lambda_i$ is the corresponding eigenvalue, we have $\bm{{{\rm L}}} {\bm{{\rm u}}}_i=\lambda_i {\bm{{\rm u}}}_i$. Since $\bm{{{\rm L}}}^n {\bm{{\rm u}}}_i=\lambda_i^n {\bm{{\rm u}}}_i$, $\bm{{{\rm u}}}_i$ is also the eigenvector of $\bm{{{\rm L}}}^n$, therefore $\bm{{{\rm F}}}$ contains information of all-order connections in the hypergraph. Constructed with the spectral features, latent communities/categories also take the high-order information into consideration, that is the superiority of our model over existing learning to rank methods.

\section{Experiments}
In this section, we design experiments on real-world datasets to validate the effectiveness of our models. We evaluate our proposed methods focusing on the following three key research questions:

\vspace{2mm}
\noindent \textbf{RQ1:} How is the performance of our entire spectrum-enhanced recommendation model (effectiveness of SCF\_SPLR)?

\noindent \textbf{RQ2:} How is the performance of the spectral features to model users' preference and items' properties (effectiveness of SCF\_BPR)?

\noindent \textbf{RQ3:} How is the performance of the SPLR optimization criterion (effectiveness of MF\_SPLR)? 
\vspace{2mm}

\subsection{Datasets} 
\begin{table}[ht!]
    \caption{Statistics of datasets}
    \centering
    \label{tab:datasets}
    \scalebox{1}{
    \begin{tabular}{ccccc}
        \hline
        Dataset & Purchase & User & Item & Sparsity\\
        \hline
        \textit{Clothes} & 115841 & 32728 & 8777 & 99.9597\% \\
        \textit{Jewelry} & 37314 & 15924 & 3607 & 99.9350\% \\
        \hline
    \end{tabular}}
\end{table}
In this paper, we perform experiments on the \emph{Amazon} dataset \cite{VBPR}, which is the consumption records from \emph{Amazon.com}. We consider two large categories \emph{Amazon.clothes} and \emph{Amazon.jewelry}, which are filtered from the \emph{Amazon} dataset. We take users' review histories as implicit feedback. Some details of the datasets are shown in Table \ref{tab:datasets}.

\begin{table*}[t]
\caption{Recommendation performance (test set)}  
\begin{center}  
\label{tab:exper_result}
\scalebox{1}{
\begin{tabular}{m{1.1cm}<{\centering}||m{1.3cm}<{\centering}|c|cccccccc}  
\hline
\hline
\multirow{2}{*}{Datasets} & \multicolumn{2}{c|}{\multirow{2}{*}{Metrics (\%)}} & \multirow{2}{*}{MP} & \multirow{2}{*}{PMF} & \multirow{2}{*}{BPR} & \multirow{2}{*}{GBPR} & \multirow{2}{*}{VBPR} & \multirow{2}{*}{SPLR} & \multicolumn{2}{c}{Improvement} \\
 & \multicolumn{2}{c|}{} & & & & & & & BPR & GBPR\\
\hline
\hline

\multirow{8}{*}{\textit{Jewelry}} & \multirow{4}{*}{$F$-1@} & 2 & $0.808{\scriptstyle\pm0.048}$ & $3.851{\scriptstyle\pm0.144}$ & $3.684{\scriptstyle\pm0.194}$ & $3.811{\scriptstyle\pm0.178}$ & $3.998{\scriptstyle\pm0.165}$ & ${\bm{4.104}\scriptstyle\pm0.185}$ &$11.40\%$ & $7.69\%$\\

 & & 5 & $0.747{\scriptstyle\pm0.039}$ & $3.507{\scriptstyle\pm0.120}$ & $3.576{\scriptstyle\pm0.137}$ & $3.656{\scriptstyle\pm0.115}$ & $3.820{\scriptstyle\pm0.068}$ & $\bm{3.905}{\scriptstyle\pm0.136}$ & $9.20\%$ & $6.81\%$ \\
 
 & & 10 & $0.608{\scriptstyle\pm0.012}$ & $3.064{\scriptstyle\pm0.082}$ & $3.159{\scriptstyle\pm0.099}$ & $3.192{\scriptstyle\pm0.102}$ & $\bm{3.354}{\scriptstyle\pm0.109}$ & $3.339{\scriptstyle\pm0.068}$ & $5.70\%$ & $4.61\%$ \\
 
 & & 20 & $0.425{\scriptstyle\pm0.014}$ & $2.403{\scriptstyle\pm0.026}$ & $2.489{\scriptstyle\pm0.042}$ & $2.515{\scriptstyle\pm0.040}$ & $\bm{2.653}{\scriptstyle\pm0.046}$ & $2.638{\scriptstyle\pm0.064}$ & $5.99\%$ & $4.89\%$ \\
 
\cline{2-11}    
 & \multirow{4}{*}{$NDCG$@} & 2 & $0.618{\scriptstyle\pm0.047}$ & $3.762{\scriptstyle\pm0.128}$ & $3.722{\scriptstyle\pm0.184}$ & $3.845{\scriptstyle\pm0.185}$ & $3.960{\scriptstyle\pm0.192}$ & $\bm{4.057}{\scriptstyle\pm0.179}$ & $9.00\%$ & $5.51\%$ \\
 
 & & 5 & $0.513{\scriptstyle\pm0.014}$ & $2.827{\scriptstyle\pm0.075}$ & $2.775{\scriptstyle\pm0.106}$ & $2.924{\scriptstyle\pm0.104}$ & $3.075{\scriptstyle\pm0.102}$ & $\bm{3.099}{\scriptstyle\pm0.121}$ & $11.68\%$ & $5.98\%$ \\

 & & 10 & $0.402{\scriptstyle\pm0.012}$ & $2.229{\scriptstyle\pm0.047}$ & $2.227{\scriptstyle\pm0.039}$ & $2.301{\scriptstyle\pm0.047}$ & $2.445{\scriptstyle\pm0.043}$ & $\bm{2.450}{\scriptstyle\pm0.048}$ & $10.01\%$ & $6.48\%$ \\

 & & 20 & $0.280{\scriptstyle\pm0.009}$ & $1.735{\scriptstyle\pm0.039}$ & $1.738{\scriptstyle\pm0.017}$ & $1.788{\scriptstyle\pm0.036}$ & $\bm{1.892}{\scriptstyle\pm0.043}$ & $1.861{\scriptstyle\pm0.046}$ & $7.08\%$ & $4.08\%$ \\

\hline
\hline
\multirow{8}{*}{\textit{Clothes}} & \multirow{4}{*}{$F$-1@} & 2 & $0.601{\scriptstyle\pm0.043}$ & $2.638{\scriptstyle\pm0.063}$ & $2.831{\scriptstyle\pm0.116}$ & $2.918{\scriptstyle\pm0.107}$ & $3.078{\scriptstyle\pm0.098}$ & $\bm{3.125}{\scriptstyle\pm0.087}$ & $10.39\%$ & $7.09\%$ \\

 & & 5 & $0.563{\scriptstyle\pm0.032}$ & $2.481{\scriptstyle\pm0.140}$ & $2.689{\scriptstyle\pm0.041}$ & $2.803{\scriptstyle\pm0.045}$ & $2.945{\scriptstyle\pm0.048}$ & $\bm{2.999}{\scriptstyle\pm0.070}$ & $11.53\%$ & $6.99\%$ \\
 
 & & 10 & $0.559{\scriptstyle\pm0.021}$ & $2.087{\scriptstyle\pm0.131}$ & $2.322{\scriptstyle\pm0.039}$ & $2.445{\scriptstyle\pm0.040}$ & $2.464{\scriptstyle\pm0.040}$ & $\bm{2.531}{\scriptstyle\pm0.063}$ & $9.00\%$ & $3.52\%$ \\
 
 & & 20 & $0.462{\scriptstyle\pm0.027}$ & $1.626{\scriptstyle\pm0.065}$ & $1.828{\scriptstyle\pm0.056}$ & $1.914{\scriptstyle\pm0.054}$ & $1.957{\scriptstyle\pm0.053}$ & $\bm{2.034}{\scriptstyle\pm0.022}$ & $11.27\%$ & $6.27\%$ \\

\cline{2-11}
 & \multirow{4}{*}{$NDCG$@} & 2 & $0.619{\scriptstyle\pm0.032}$ & $2.791{\scriptstyle\pm0.052}$ & $3.026{\scriptstyle\pm0.134}$ & $3.118{\scriptstyle\pm0.126}$ & $3.119{\scriptstyle\pm0.118}$ & $\bm{3.331}{\scriptstyle\pm0.144}$ & $10.08\%$ & $6.83\%$ \\
 
 & & 5 & $0.452{\scriptstyle\pm0.039}$ & $2.021{\scriptstyle\pm0.109}$ & $2.185{\scriptstyle\pm0.068}$ & $2.293{\scriptstyle\pm0.054}$ & $2.410{\scriptstyle\pm0.040}$ & $\bm{2.451}{\scriptstyle\pm0.028}$ & $12.17\%$ & $6.89\%$ \\
 
 & & 10 & $0.373{\scriptstyle\pm0.009}$ & $1.567{\scriptstyle\pm0.091}$ & $1.740{\scriptstyle\pm0.052}$ & $1.844{\scriptstyle\pm0.047}$ & $1.865{\scriptstyle\pm0.042}$ & $\bm{1.907}{\scriptstyle\pm0.042}$ & $9.60\%$ & $3.42\%$ \\
 
 & & 20 & $0.300{\scriptstyle\pm0.007}$ & $1.197{\scriptstyle\pm0.054}$ & $1.330{\scriptstyle\pm0.069}$ & $1.402{\scriptstyle\pm0.045}$ & $1.442{\scriptstyle\pm0.022}$ & $\bm{1.477}{\scriptstyle\pm0.020}$ & $11.05\%$ & $5.35\%$ \\
 
\hline
\hline
\end{tabular}} 
\end{center} 
\end{table*}  

\subsection{Baselines}
We adopt the following methods as baselines for performance comparison: 
\begin{itemize}
\item{\textbf{MP:} This \textbf{M}ost \textbf{P}opular method ranks items according to their popularity. It is a non-personalized method to benchmark the recommendation performances.}

\item{\textbf{PMF:} This \textbf{P}robabilistic \textbf{M}atrix \textbf{F}actorization method, which was proposed by \cite{O3}, is a frequently used state-of-the-art approach for rating-based optimization and prediction. We set the score of positive samples as 1 and missing values as 0.}

\item{\textbf{BPR (MF\_BPR):} This \textbf{B}ayesian \textbf{P}ersonalized \textbf{R}anking me-thod is the most widely used ranking-based method for implicit feedback \cite{BPR}. It regards all unobserved samples as negative samples and maximizes the likelihood of users' preference over a pair of positive sample and negative sample.}

\item{\textbf{GBPR (MF\_GBPR):} This \textbf{G}roup Preference-based \textbf{B}ayesian \textbf{P}ersonalized \textbf{R}anking method \cite{groupBPR} is an extension of BPR, which tries to relax BPR's assumptions to a group pairwise preference assumption. We fix the number of grouped users to 3.}

\item{\textbf{VBPR:} This \textbf{V}isual \textbf{B}ayesian \textbf{P}ersonalized \textbf{R}anking method is a stat-of-the-art visual-based recommendation method \cite{VBPR}. The visual features are extracted from the product images with a pre-trained deep convolutional neural network (CNN).}
\end{itemize}

\subsection{Experiment Settings}
The eigen-decomposition of Laplacian matrices is implemented with \texttt{sparse.linalg()} function of \texttt{scipy} library in \texttt{python}. The \textit{Amazon} dataset is filtered with 5-core (remove users and items with less than 5 purchase records) and records before 2010 is removed. We then filter \textit{Jewelry} and \textit{Clothes} datasets from \textit{Amazon}. Each dataset is split into training set (80\%), validation set(10\%), and test set (10\%) randomly. Cold items and users (items and users with no record in training set) in validation and test sets are removed. We adapt $F_1$-score and normalized discounted cumulative gain ($NDCG$) to evaluate the performance of the baselines and our proposed model. Mini-batch stochastic gradient descent (MSGD) is leveraged to optimize our model. The learning rate is determined by grid searching in the range of $\{0.005,$ $0.01,$ $0.02,$ $0.05,$ $0.1,$ $0.2\}$ and the batch size is determined in the range of $\{1000,$ $2000,$ $3000,$ $4000,$ $5000\}$. We evaluate different number of latent factors $K_0$ in the range of $\{10,$ $20,$ $50,$ $100,$ $200\}$, the regularization coefficient $\lambda_r$ in the range of $\{0,$ $0.1,$ $0.2,$ $\cdots,$ $1.5\}$, and the weighted parameters $\eta_1$ and $\eta_2$ in the range of $\{0, 0.01, 0.1\}$. We conduct our experiments by predicting top-$\{2,$ $5,$ $10,$ $20\}$ items to each user. The sampling rate $m$ is set as 5 (select 5 negative/potential samples for each positive sample to construct pairs) in all pairwise learning to balance the accuracy and efficiency.

\subsection{Performance of Our Entire Model (RQ1)}
In this subsection, we report the performance of our entire model (SCF model optimized with SPLR, SCF\_SPLR, abbreviated as SPLR) and baselines to give some analysis. We tune all models in the validation set and test the performance in the test set. The impact of some important hyperparameters, such as the number of latent factors, regularization coefficient, and weighted parameters, are shown. When training, we iterate the whole dataset 200 times to learn models and select 1000 samples randomly from the test/validation set to test all models in each iteration (except MP, we just test 200 times without training). We record the best performance of each model during this procedure as the evaluation of it. We execute all models 5 times and then report the mean and standard deviation. The learning rate is set as 0.05 and the batch size is set as 5000 for all models. 

The performance of all models is represented in Table \ref{tab:exper_result}, we can see that all personalized models outperform MP significantly in all situations --- about 3-6 times improvement. Comparison between BPR and PMF shows the superiority of pairwise learning over point-wise one. Considering the group behaviour, GBPR gets a further enhancement and outperforms BPR in all situations. As an enhanced learning to rank method, we compare our model with BPR and GBPR, and the improvement is shown in the last two columns in Table \ref{tab:exper_result}. We can see that our model dramatically outperforms these two models. In GBPR, only one-order user connections are utilized while in SPLR, the spectral features capture high-order connections for both users and items, thus provides high-level and comprehensive description of the similarity among vertices.

We also report the performance of VBPR, the state-of-the-art side information-based model. With the extra visual feature, it performs the best among all baselines. As we can see, our model can even outperform VBPR in most situations. Though both utilizing ``side'' information to enhance the accuracy, VBPR uses ``outside'' information, i.e., extra visual data, while SPLR uses ``inside'' information extracted from the purchase records. From the last two columns of Table \ref{tab:exper_result}, we can see that in \textit{Jewelry} dataset, the relative improvement of our model decreases with the increasing of $n$ (the number to recommend), while in \textit{Clothes} dataset, the relative improvement is stable. That is not because the our model performs badly with a large $n$, it is the property of the dataset. We can see that in \textit{Jewelry} set, the gap (relative improvement) between any two pairwise learning models decreases with the increasing of $n$.

\begin{figure}[ht!]
\setlength{\abovecaptionskip}{2mm}
\centering
\subfigure[\textit{Jewelry}]{
\includegraphics[scale = 0.37]{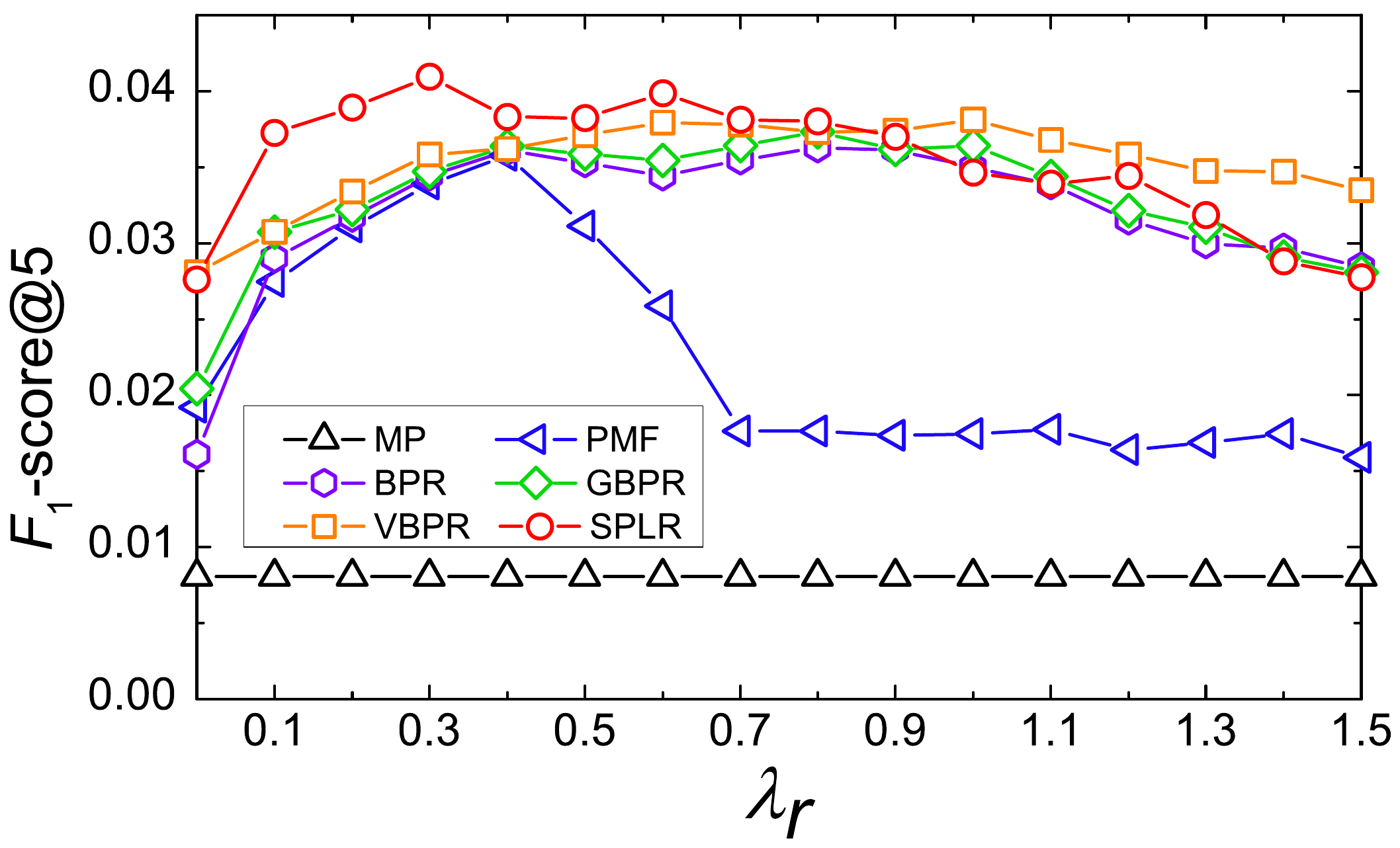}
\label{fig:lambda_r_jewelry}
}
\subfigure[\textit{Clothes}]{
\includegraphics[scale = 0.37]{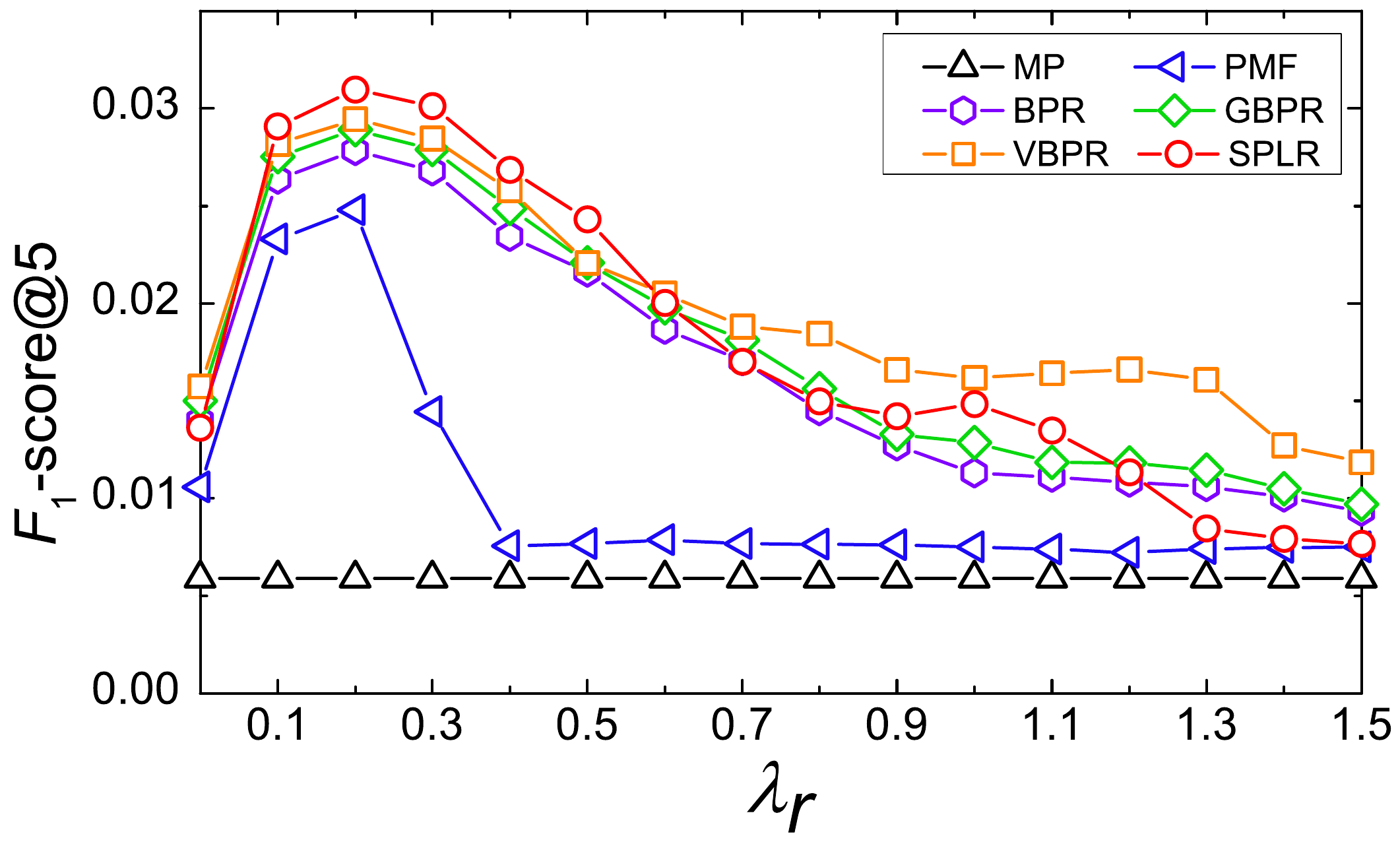}
\label{fig:lambda_r_clothes}
}
\caption{Impact of regularization coefficient $\lambda_r$ (validation set)}
\label{fig:lambda_r}
\end{figure}

To analyze the sensitivity of our model, we report the performance ($F_1$-score@5) with different hyperparameters. The impact of regularization coefficient $\lambda_r$ is represented in Figure \ref{fig:lambda_r}. Though both filtered from \textit{Amazon} dataset, these two datasets have pretty different properties. From Figure \ref{fig:lambda_r_jewelry}, we can see that all models perform quiet differently to each other in \textit{Jewelry} dataset, PMF, BPR, GBPR, VBPR, and SPLR get the best performance when $\lambda_r$ is set to 0.4, 0.8, 0.8, 1.0, and 0.3 respectively. While in Figure \ref{fig:lambda_r_clothes}, all models get the best performance when $\lambda_r$ is set to 0.2. With the increasing of $\lambda_r$, accuracy of the point-wise learning method decreases rapidly compared with pairwise learning methods.

\begin{figure}[ht!]
\setlength{\abovecaptionskip}{2mm}
\centering
\subfigure[\textit{Jewelry}]{
\includegraphics[scale = 0.27]{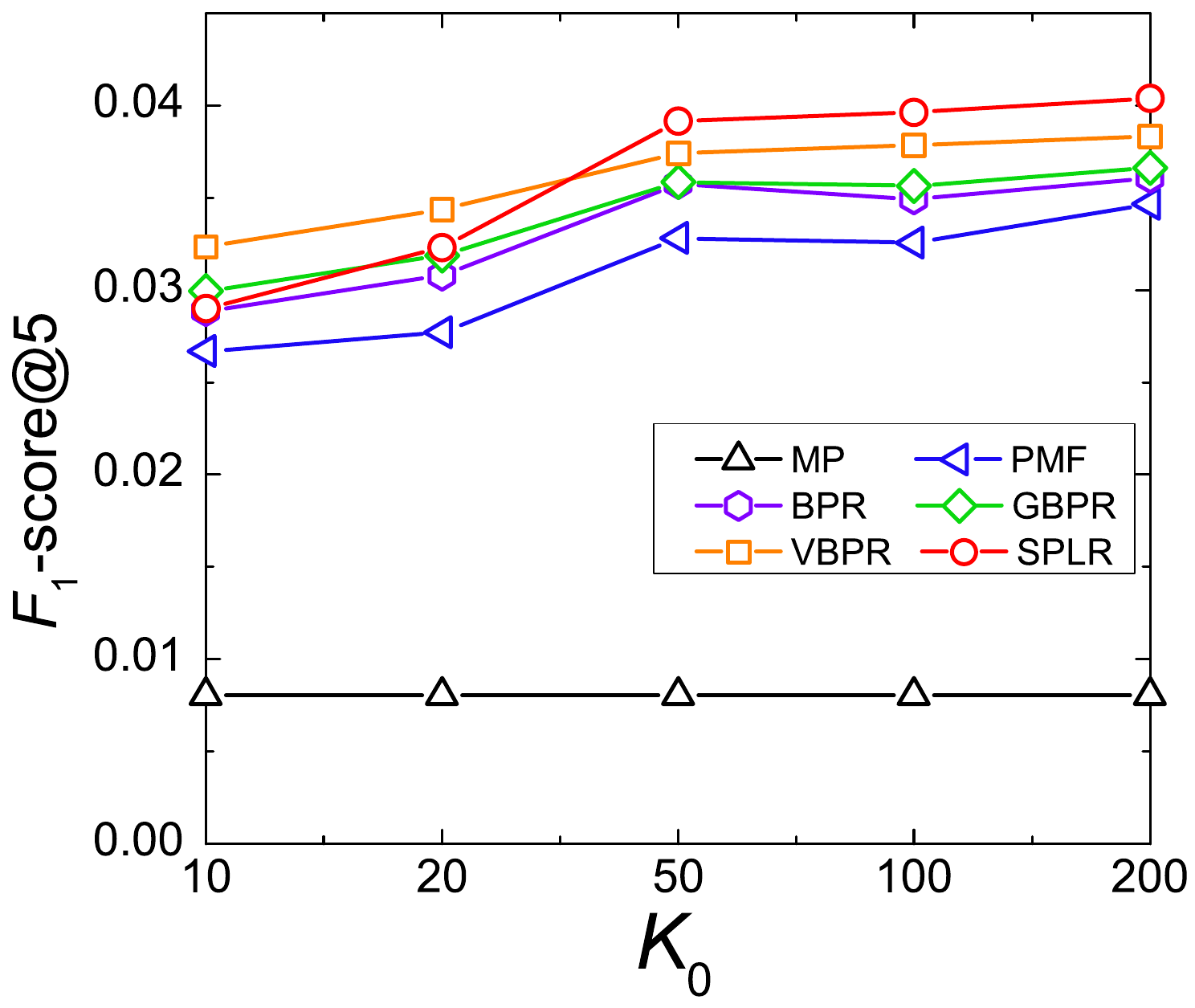}
\label{fig:K_jewelry}
}
\hspace{-3mm}
\subfigure[\textit{Clothes}]{
\includegraphics[scale = 0.27]{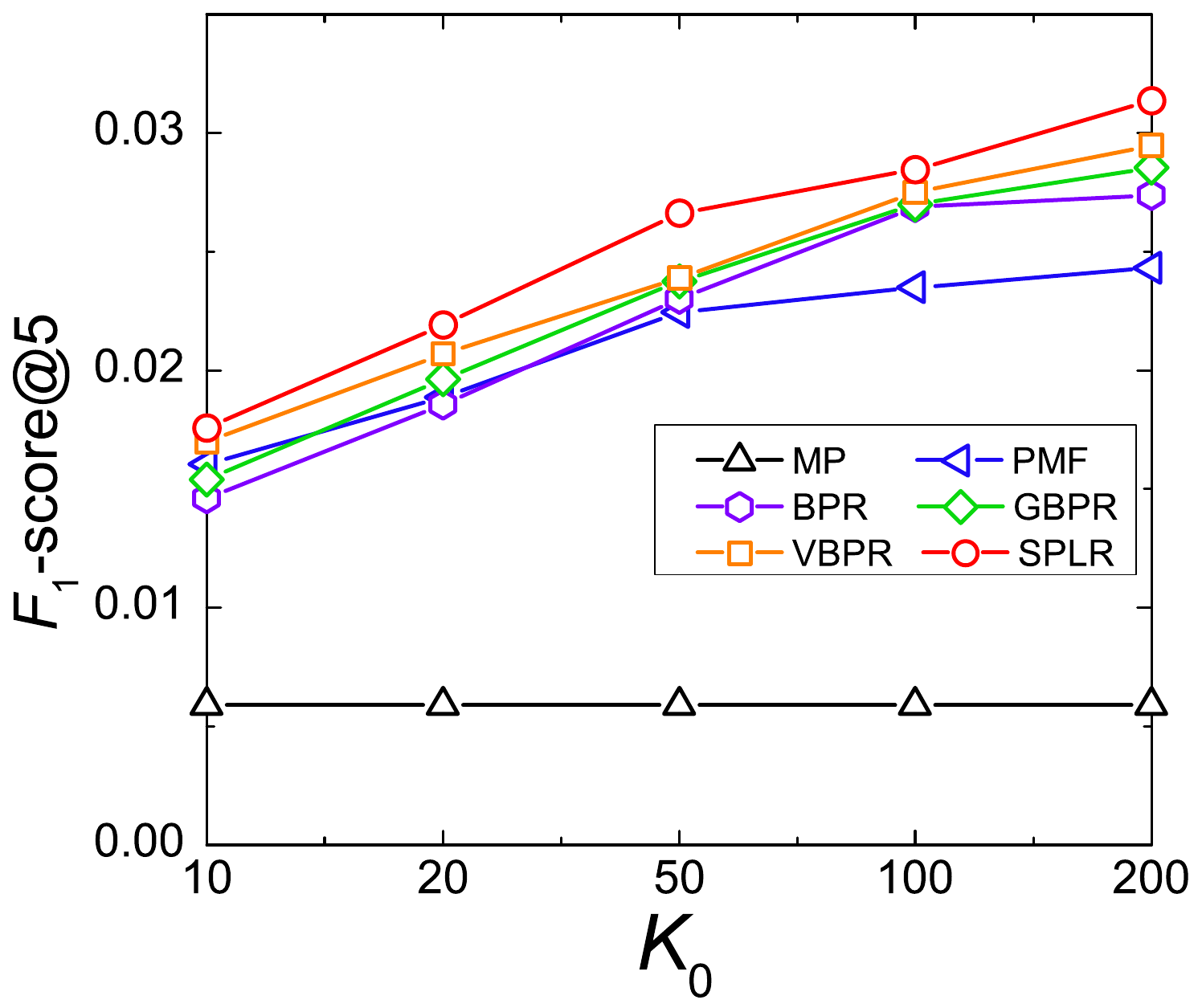}
\label{fig:K_clothes}
}
\caption{Impact of latent factor dimensions $K_0$ (validation set)}
\label{fig:K}
\end{figure}

We also report $F_1$-score@5 with different length of latent factors to get the best $K_0$. In both two datasets, performance of all models (except MP) increase with the increasing of dimensions. We can see that due to the smaller size, the purchase record matrix of \textit{Jewelry} has a smaller rank than that of \textit{Clothes}: The performance in \textit{Jewelry} tends to be stable after $K_0$ is larger than 50 while the performance in \textit{Clothes} keeps increasing obviously with $K_0$ changing from 10 to 200. To get the best performance, we set $K_0$ as 200 for all models.

\begin{figure}[ht!]
\setlength{\abovecaptionskip}{2mm}
\centering
\subfigure[\textit{Jewelry}]{
\includegraphics[scale = 0.17]{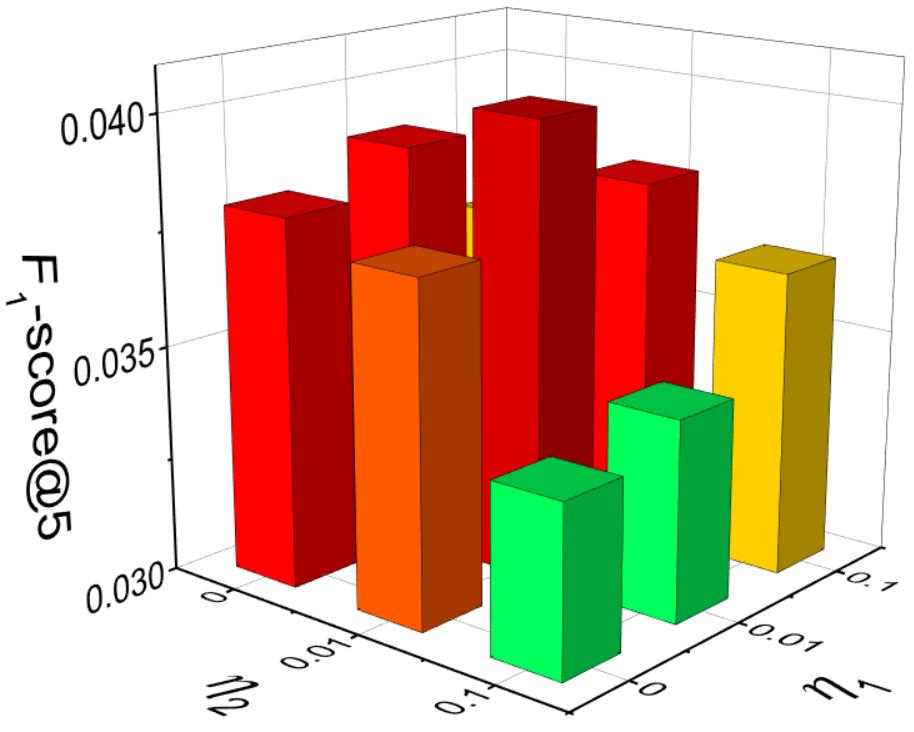}
\label{fig:eta_jewelry}
}
\hspace{-3mm}
\subfigure[\textit{Clothes}]{
\includegraphics[scale = 0.17]{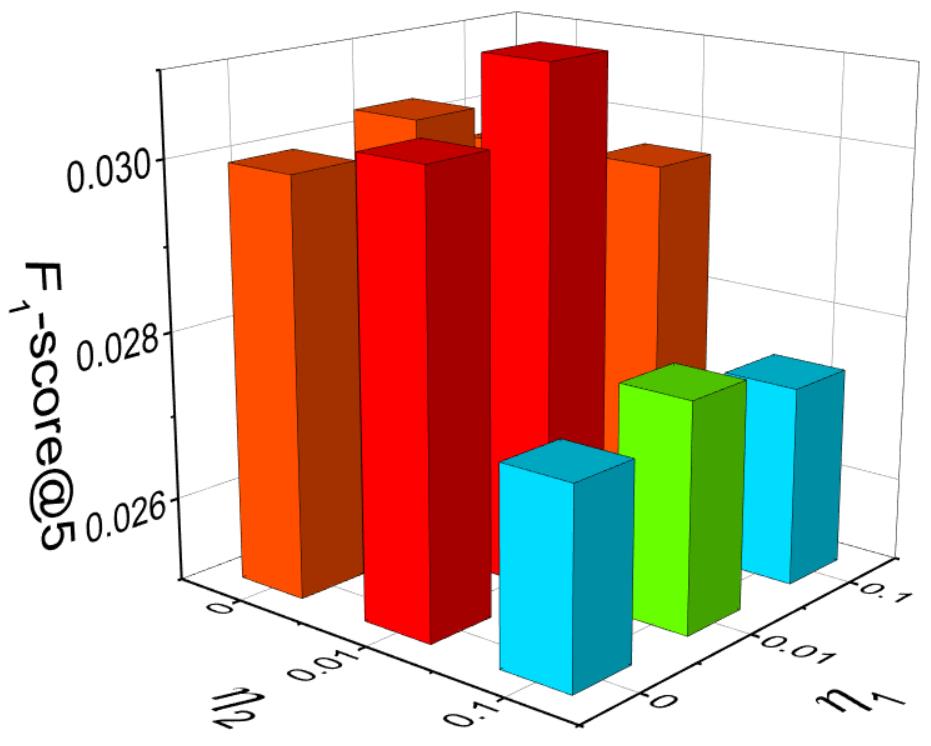}
\label{fig:eta_clothes}
}
\caption{Impact of weighting parameters $\eta_1$ and $\eta_2$ (validation set)}
\label{fig:eta}
\end{figure}

The impact of weighting parameters is illustrated in Figure \ref{fig:eta}. Our model performs the best when $\eta_1=0.01$ and $\eta_2=0.01$ in these two datasets. When $\eta_1=0$ and $\eta_2=0$, SPLR (SCF\_SPLR) degenerates into SCF\_BPR. From the figures we can see that SPLR outperforms SCF\_BPR 4.70\% in \textit{Jewelry} validation dataset and 4.03\% in \textit{Jewelry} validation dataset on $F_1$-score@5. We can see that the enhancement of SLPR is not so significant, this is because we have explored the similarity information by incorporating the spectral features into SCF\_BPR model, which outperforms MF\_BPR significantly (Table \ref{tab:exper_result2}). We leverage SPLR optimization to explore the similarity information thoroughly for further accuracy improvement. 

We demonstrated the effectiveness of our entire model in this subsection, and in next subsections, we illustrate the effectiveness of each part of our model --- spectral feature-enhanced model (SCF\_BPR) and spectral clustering-enhanced pairwise learning optimization strategy (MF\_SPLR).

\subsection{Effectiveness of SCF\_BPR (RQ2)}

In this subsection, we denote all models with the model names and optimization methods. We rename BPR as MF\_BPR, which means MF model optimized with BPR algorithm. Similarly, SCF\_BPR is SCF model optimized with BPR. We demonstrate the effectiveness of spectral features by comparing SCF\_BPR with other baselines. We first report the sensitivity with the number of eigenvectors by grid searching in the range of \{0,10,100,1000\}. All experiments in this and next subsection are conducted in \textit{Jewelry} dataset.

\begin{figure}[ht!]
\includegraphics[scale = 0.25]{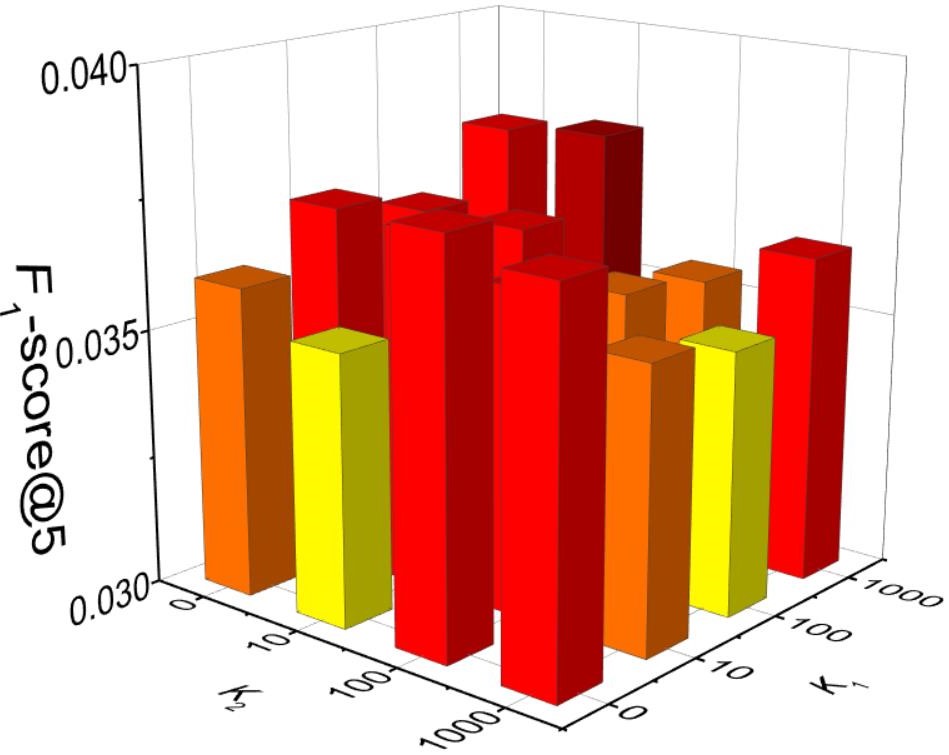}
\caption{Performance of SCF\_BPR with different length of spectral features (\textit{Jewelry}, validation set)}
\label{fig:len_spe_fea}
\end{figure}

The sensitivity of SCF\_BPR with the length of spectral features is shown in Figure \ref{fig:len_spe_fea}. When $K_1=1000$ and $K_2=10$, SCF\_BPR performs the best. When $K_1=0$ and $K_2=0$, SCF\_BPR degenerates into MF\_BPR. SCF\_BPR outperforms MF\_BPR 7.2\% on $F_1$-score@5 in \textit{Jewelry} validation set. From Figure \ref{fig:len_spe_fea} we can see that user feature and item feature do not jointly work well since SCF\_BPR does not perform well when $K_1$ and $K_2$ are both large. In real-world application, we can set $K_1=0$ and $K_2=100$ to balance the accuracy and the efficiency.

\begin{table}[ht!]
\caption{Performance comparison for SCF\_BPR (\textit{Jewelry}, test set)}  
\begin{center}  
\label{tab:exper_result2}
\scalebox{1}{
\begin{tabular}{m{1.1cm}<{\centering}|c|ccc}  
\hline
\hline
\multicolumn{2}{c|}{Metrics (\%)} & MF\_BPR & SCF\_BPR & Improvement \\
\hline
\hline

\multirow{4}{*}{$F$-1@} & 2 & $3.684{\scriptstyle\pm0.194}$ & ${\bm{3.952}\scriptstyle\pm0.221}$ & $7.27\%$\\

 & 5 & $3.576{\scriptstyle\pm0.137}$ & $\bm{3.768}{\scriptstyle\pm0.098}$ & $5.37\%$ \\
 
 & 10 & $3.159{\scriptstyle\pm0.099}$ & $\bm{3.301}{\scriptstyle\pm0.087}$ & $4.50\%$ \\
 
 & 20 & $2.489{\scriptstyle\pm0.042}$ & $\bm{2.596}{\scriptstyle\pm0.029}$ & $4.30\%$ \\
 
\cline{1-5}    
\multirow{4}{*}{$NDCG$@} & 2 & $3.722{\scriptstyle\pm0.184}$ & $\bm{3.953}{\scriptstyle\pm0.151}$ & $6.21\%$ \\
 
 & 5 & $2.775{\scriptstyle\pm0.106}$ & $\bm{2.952}{\scriptstyle\pm0.074}$ & $6.38\%$ \\

 & 10 & $2.227{\scriptstyle\pm0.039}$ & $\bm{2.366}{\scriptstyle\pm0.037}$ & $6.24\%$ \\

 & 20 & $1.738{\scriptstyle\pm0.017}$ & $\bm{1.835}{\scriptstyle\pm0.023}$ & $5.58\%$ \\
 
\hline
\hline
\end{tabular}} 
\end{center} 
\end{table}

The performance of SCF\_BPR and baselines in test set is shown in Table \ref{tab:exper_result2}. To save space, only the performance of MF\_BPR is reported, since it is the most important baseline. We can check Table \ref{tab:exper_result} for the performance of other baselines. With the spectral features providing similarity information, SCF\_BPR outperforms MF\_BPR 7.27\% and 6.38\% on $F_1$-score and $NDCG$ respectively in the best situation.

\subsection{Effectiveness of MF\_SPLR (RQ3)}

\begin{figure}[ht!]
\includegraphics[scale = 0.25]{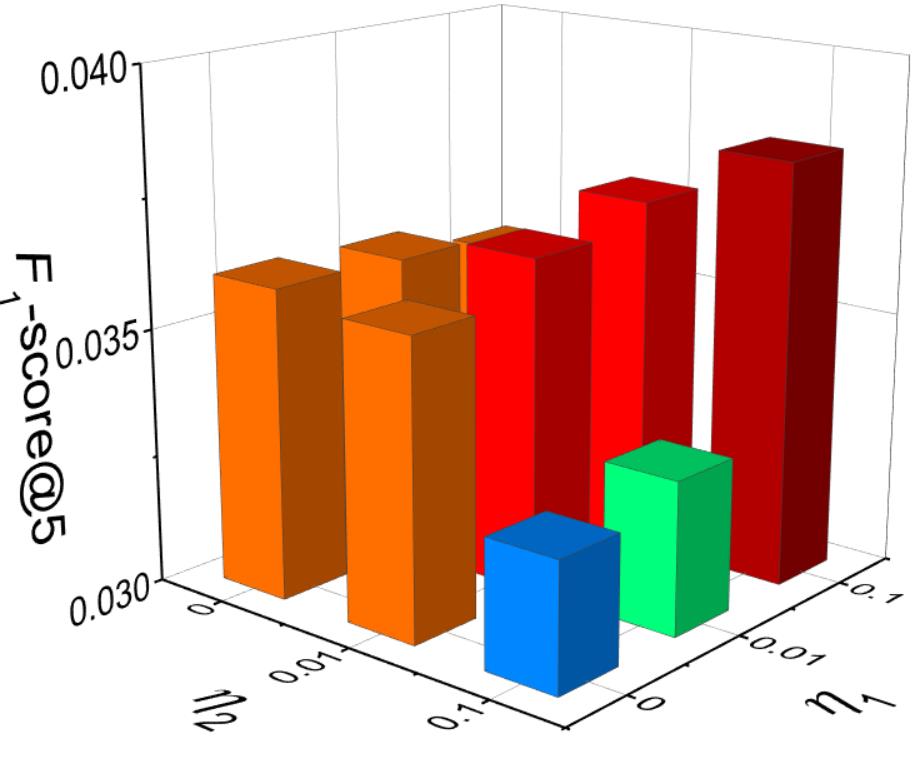}
\caption{Performance of MF\_SPLR with different weighting parameters (\textit{Jewelry}, validation set)}
\label{fig:eta1_eta2}
\end{figure}

The sensitivity of MF\_SPLR with the weighting parameters $\eta_1$ and $\eta_2$ are shown in Figure \ref{fig:eta1_eta2}. We can see that MF\_SPLR performs the best when $\eta_1 = 0.1$ and $\eta_2 = 0.1$. When $\eta_1 = 0$ and $\eta_2 = 0$, MF\_SPLR degenerates to MF\_BPR. MF\_SPLR outperforms MF\_BPR 6.13\% $F_1$-score@5 in \textit{Jewelry} validation set. Comparing Figure \ref{fig:eta_jewelry} and Figure \ref{fig:eta1_eta2} we can see that though both enhanced with SPLR optimization, MF\_SPLR gains more improvement than SCF\_SPLR, and the SPLR terms are weighted more in MF\_SPLR ($\eta_1=0.1$, $\eta_2=0.1$ in MF\_SPLR while $\eta_1=0.01$, $\eta_2=0.01$ in SCF\_SPLR). It is because that the similarity information has been leveraged in SCF\_SPLR by modeling with the spectral features. While in SCF\_SPLR, SPLR is the only way to utilize the similarity information. 

\begin{table}[ht!]
\caption{Performance comparison for MF\_SPLR (\textit{Jewelry}, test set)}  
\begin{center}  
\label{tab:exper_result3}
\scalebox{1}{
\begin{tabular}{m{1.1cm}<{\centering}|c|ccc}  
\hline
\hline
\multicolumn{2}{c|}{Metrics (\%)} & MF\_BPR & MF\_SPLR & Improvement \\
\hline
\hline
\multirow{4}{*}{$F$-1@} & 2 & $3.684{\scriptstyle\pm0.194}$ & ${\bm{3.986}\scriptstyle\pm0.074}$ & $8.20\%$\\
 & 5 & $3.576{\scriptstyle\pm0.137}$ & $\bm{3.789}{\scriptstyle\pm0.123}$ & $5.96\%$ \\
 & 10 & $3.159{\scriptstyle\pm0.099}$ & $\bm{3.314}{\scriptstyle\pm0.056}$ & $4.91\%$ \\
 & 20 & $2.489{\scriptstyle\pm0.042}$ & $\bm{2.578}{\scriptstyle\pm0.054}$ & $3.58\%$ \\
\cline{1-5}    
\multirow{4}{*}{$NDCG$@} & 2 & $3.722{\scriptstyle\pm0.184}$ & $\bm{3.924}{\scriptstyle\pm0.151}$ & $5.43\%$ \\
 & 5 & $2.775{\scriptstyle\pm0.106}$ & $\bm{2.923}{\scriptstyle\pm0.074}$ & $5.33\%$ \\
 & 10 & $2.227{\scriptstyle\pm0.039}$ & $\bm{2.339}{\scriptstyle\pm0.037}$ & $5.03\%$ \\
 & 20 & $1.738{\scriptstyle\pm0.017}$ & $\bm{1.797}{\scriptstyle\pm0.023}$ & $3.39\%$ \\
\hline
\hline
\end{tabular}} 
\end{center} 
\end{table} 

The performance of MF\_SPLR and baselines in \textit{Jewelry} test set is shown in Table \ref{tab:exper_result3}. From the table we can see that SCF\_BPR outperforms MF\_BPR 8.20\% and 5.42\% on $F_1$-score and $NDCG$ respectively in the best situation. Comparing Tables \ref{tab:exper_result}, \ref{tab:exper_result2}, and \ref{tab:exper_result3}, we can see that the improvement of the entire model is less than the sum of the improvement of each part. That may be because our model utilizes the similarity information by modeling (SCF) and optimizing (SPLR), thus these two parts provide the same information in different ways. 

\begin{figure*}[ht!]
\setlength{\abovecaptionskip}{2mm}
\centering
\subfigure[]{
\includegraphics[scale = 0.35]{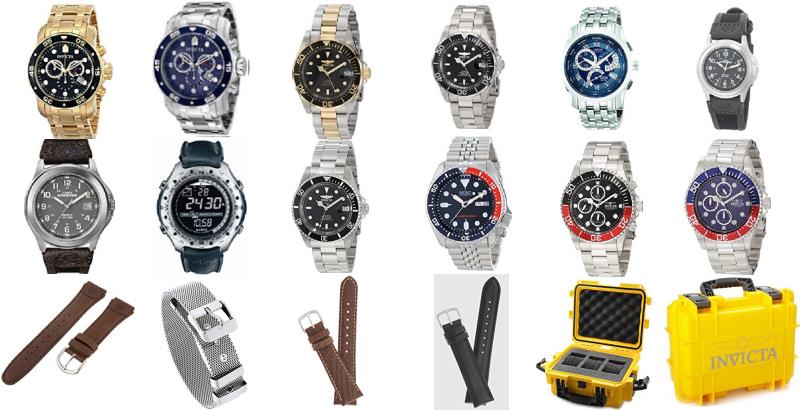}
\label{subfig:c1}
}
\hspace{1cm}
\subfigure[]{
\includegraphics[scale = 0.35]{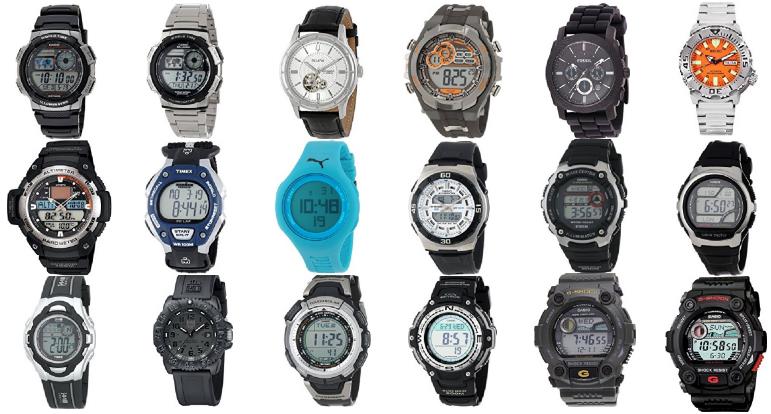}
\label{subfig:c2}
}
\subfigure[]{
\includegraphics[scale = 0.35]{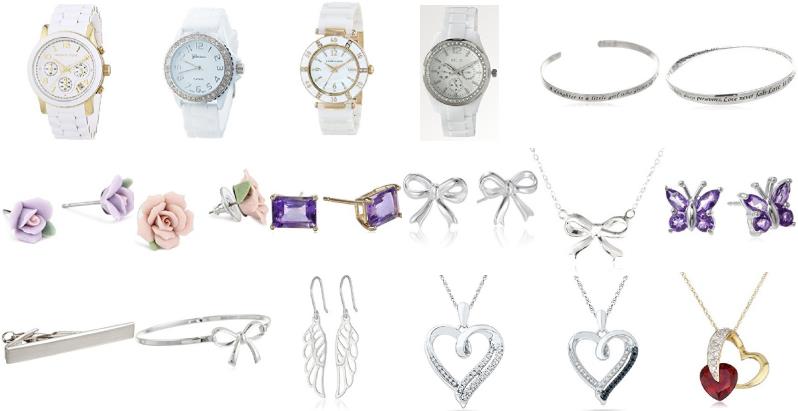}
\label{subfig:c3}
}
\hspace{1cm}
\subfigure[]{
\includegraphics[scale = 0.35]{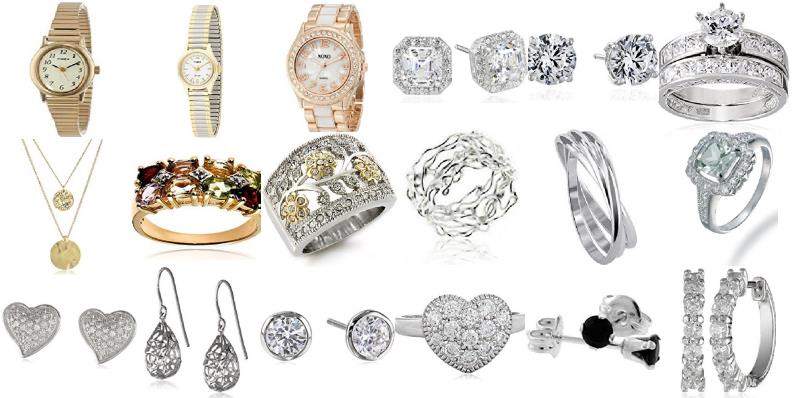}
\label{subfig:c4}
}
\subfigure[]{
\includegraphics[scale = 0.35]{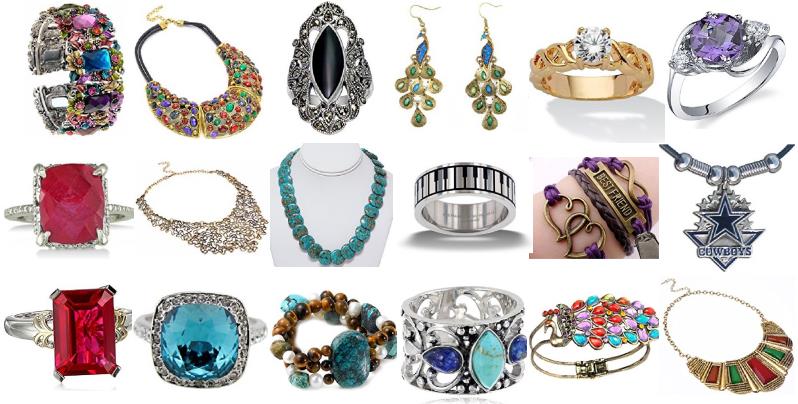}
\label{subfig:c5}
}
\hspace{1cm}
\subfigure[]{
\includegraphics[scale = 0.35]{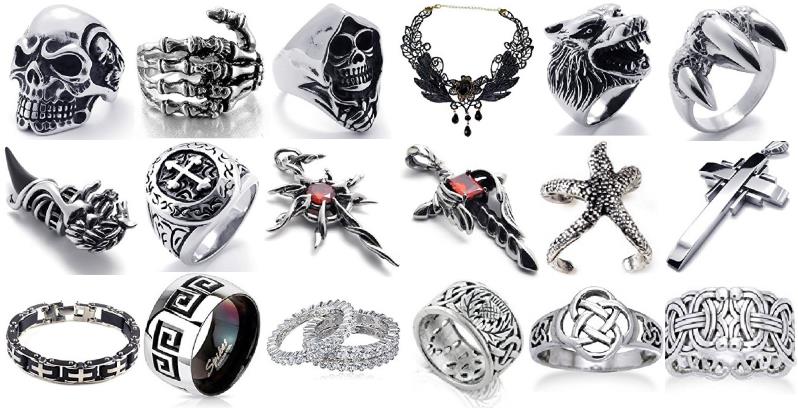}
\label{subfig:c6}
}
\caption{Latent categories (\textit{Jewelry} set)}
\label{fig:latent_categories}
\end{figure*}

In Figure \ref{fig:latent_categories}, we show some latent categories of the \textit{Jewelry} set, which contains all watches and jewelries. There are six latent categories in Figure \ref{fig:latent_categories}. Not like a real category, items in a latent category are relevant items or similar items (items strongly connected in the hypergraph). For example, in Figure \ref{subfig:c1}, items are watches and some relevant commodities, such as watch bands and watch boxes\footnote{\url{https://www.amazon.com/dp/B005IHDLYC/?tag=tc0f3f-20}}. Users who want to by watches may have interests in this category. Though items in the same latent categories may be different kinds of commodities, they are in the same style. For example, items in Figure \ref{subfig:c1} are designed for business men, they are luxurious, with metallic luster, look mature and steady. On the contrary, different categories may contain items of the same kind, but they are in different styles. For example, the category in Figure \ref{subfig:c2} is also for watches, however is in a totally different style from that in Figure \ref{subfig:c1}. Most items in this category are plastic and cheap digital watches for sports and boys may have interests in them. Items in Figure \ref{subfig:c3} are watches and jewelries for young girls, they look simple and elegant. Jewelries in Figure \ref{subfig:c4} and Figure \ref{subfig:c5} are luxurious and exaggerative, which are full of diamonds and gemstones. Items in Figure \ref{subfig:c6} are in punk style and rebellious teenagers may prefer them.

Figure \ref{fig:latent_categories} shows how the item spectral feature helps to recommend. The item spectral feature indicates the style of the item, circle of interest the item belongs to, the price level, target users, etc. Modeling with the item spectral feature (SCF) makes the items in the same latent category tend to have similar score. And optimizing with the item spectral feature (SPLR) makes the items in the positive category (latent category containing positive samples) tend to have a high score. For the user spectral feature, we can draw the same conclusion.

Both spectral features and latent factors are extracted from graph structures which are constructed from the purchase records (one bipartite graph and two hypergraphs). In fact, they both mine the similarity information while with different emphasis --- user/item spectral features describe the similarity of users/items, while latent factors describe the similarity between a user and an item. Spectral features and latent factors complete each other in describing the similarity information and enhance the effectiveness of the recommendation model.

\section{Extension}

\begin{table*}[ht!]
\caption{Performance with different features (\textit{Jewelry}, test set)}  
\begin{center}  
\label{tab:exper_result4}
\scalebox{1}{
\begin{tabular}{m{2.2cm}<{\centering}|c|m{2.2cm}<{\centering}m{2.2cm}<{\centering}m{2.2cm}<{\centering}m{2.2cm}<{\centering}}  
\hline
\hline
\multicolumn{2}{c|}{\multirow{2}{*}{\diagbox{Metrics (\%)}{Features}}} & \multirow{2}{*}{None} & \multirow{2}{*}{Spectral features} & \multirow{2}{*}{CNN feature} & Spectral features \\
\multicolumn{2}{c|}{} & & & & \& CNN feature \\
\hline
\hline

\multirow{4}{*}{$F$-1@} & 2 & $3.684{\scriptstyle\pm0.194}$ & $4.104{\scriptstyle\pm0.185}$ & $4.218{\scriptstyle\pm0.213}$ & ${\bm{4.414}\scriptstyle\pm0.177}$ \\
 & 5 & $3.576{\scriptstyle\pm0.137}$ & $3.905{\scriptstyle\pm0.136}$ & $3.974{\scriptstyle\pm0.106}$ & $\bm{4.108}{\scriptstyle\pm0.165}$ \\
 & 10 & $3.159{\scriptstyle\pm0.099}$ & $3.339{\scriptstyle\pm0.068}$ & $3.398{\scriptstyle\pm0.103}$ & $\bm{3.567}{\scriptstyle\pm0.120}$ \\
 & 20 & $2.489{\scriptstyle\pm0.042}$ & $2.638{\scriptstyle\pm0.064}$ & $2.631{\scriptstyle\pm0.067}$ & $\bm{2.800}{\scriptstyle\pm0.103}$ \\
\cline{1-6}
\multirow{4}{*}{$NDCG$@} & 2 & $3.722{\scriptstyle\pm0.184}$ & $4.057{\scriptstyle\pm0.179}$ & $4.184{\scriptstyle\pm0.165}$ & $\bm{4.392}{\scriptstyle\pm0.278}$ \\
 & 5 & $2.775{\scriptstyle\pm0.106}$ & $3.099{\scriptstyle\pm0.121}$ & $3.173{\scriptstyle\pm0.090}$ & $\bm{3.315}{\scriptstyle\pm0.183}$ \\
 & 10 & $2.227{\scriptstyle\pm0.039}$ & $2.450{\scriptstyle\pm0.048}$ & $2.542{\scriptstyle\pm0.078}$ & $\bm{2.623}{\scriptstyle\pm0.137}$ \\
 & 20 & $1.738{\scriptstyle\pm0.017}$ & $1.861{\scriptstyle\pm0.046}$ & $1.934{\scriptstyle\pm0.073}$ & $\bm{2.011}{\scriptstyle\pm0.092}$ \\
\hline
\hline
\end{tabular}} 
\end{center} 
\end{table*} 

In this paper, we proposed new spectral features and incorporated them into an MF model to predict users' preference. We then introduced new spectral clustering-based pairwise learning method to optimize our model. In this section, we extend our model and propose a framework of learning to rank models enhanced with side information features. Not only the spectral features, all features can be leveraged in our framework. Assuming there are $n$ features for users ${\bm{{\rm E}}}^{(1)}$, $\cdots$, ${\bm{{\rm E}}}^{(n)}$ and $m$ features for items ${\bm{{\rm F}}}^{(1)}$, $\cdots$, ${\bm{{\rm F}}}^{(m)}$, the prediction is given by:
\begin{flalign}
\label{equ:hybrid_model1}
\hat{ {\bm{{\rm R}}} } = {\bm{{\rm U}}} {\bm{{\rm V}}}^\mathsf{T} + \sum_{j=1}^m{\bm{{\rm P}}}^{(j)} {{\bm{{\rm F}}}^{(j)}}^\mathsf{T} + \sum_{k=1}^n{\bm{{\rm E}}}^{(k)} {{\bm{{\rm Q}}}^{(k)}}^\mathsf{T}, \nonumber
\end{flalign}
where ${\bm{{\rm P}}}^{(j)}_u$ is the preference of user $u$ based on the $j$-th item feature. These item features can be high-level features like the spectral feature, CNN feature, auditory feature, or low-level features like the color histogram or tags. ${\bm{{\rm Q}}}^{(k)}_i$ is the fitness of item $i$ based on the $k$-th user feature. These user features can be high-level features like the spectral feature, or the low-level information vector containing such as age, genders, addresses, etc.

For pairwise learning, we use the latent community $\mathcal{C}_u = \bigcup_{j=1}^n \mathcal{C}^{(j)}_u$ and latent category $\mathcal{C}_i = \bigcup_{k=1}^m \mathcal{C}^{(k)}_i$ to construct the potential set in Equation (\ref{equ:three_sets}), where $\mathcal{C}^{(j)}_u$ is the latent community of user $u$ clustered by the $j$-th feature and $\mathcal{C}^{(k)}_i$ is the latent category of item $i$ clustered by the $k$-th feature.

Experimental results of our extended model are shown in Table \ref{tab:exper_result4}. These four columns are performances of our extended models with no feature (i.e., MF\_BPR), with the spectral features (i.e., SCF\_SPLR), with the CNN feature (i.e., VBPR optimized with visual clustering-enhanced pairwise learning to rank), and with the spectral features plus the CNN feature. Enhanced with the spectral features, SCF\_SPLR outperforms MF\_BPR 9.20\% on $F_1$-score@5. Leveraging appearance information, our extended model with the CNN feature outperforms MF\_BPR 11.13\% on $F_1$-score@5. Comparing Table \ref{tab:exper_result} and Tale \ref{tab:exper_result4} we can see that it also outperforms VBPR 4.03\% on $F_1$-score@5 due to the enhanced pairwise learning optimization. Our extended model with the CNN feature and spectral features outperforms MF\_BPR 14.88\% on $F_1$-score@5.

\section{Conclusion and Future Work}
In this paper, we investigated the usefulness of the spectral features in recommendation tasks. We first introduced novel spectral features, which contain similarity information, and injected them into an MF structure to model users' preference and items' properties. We then clustered the spectral features to construct the latent communities and categories for users and items respectively, and used them to enhance the pairwise learning. We finally extended our model and proposed a framework for side information-enhanced pairwise learning. Experiments on challenging real-world datasets show that our proposed methods significantly outperform state-of-the-art models.

For future work, we will investigate the effectiveness of our proposed spectral features in the setting of explicit feedback. Also, we are interested in explaining the recommendation result to users with the information of latent communities and categories. Lastly, we will use neural networks \cite{NCF,he_conver} to learn how to combine features, such as the item spectral feature and user spectral feature, or spectral features and other kinds of features.

\newpage
\bibliographystyle{ACM-Reference-Format}

\bibliography{_sample-bibliography.bbl}
\end{document}